\documentclass[%
reprint,
showpacs,
 amsmath,amssymb,
 aps,
 pre,
 showkeys,
floatfix,
longbibliography
]{revtex4-2}

\usepackage{graphicx} 
\usepackage{dcolumn} 
\usepackage{bm} 
\usepackage{booktabs}
\usepackage{array}
\usepackage[utf8]{inputenc}
\usepackage[T1]{fontenc}
\usepackage{times}
\usepackage{flafter} 
\usepackage{mathrsfs} 
\usepackage[hidelinks,urlcolor=blue, colorlinks=true,citecolor=blue, linkcolor=blue ]{hyperref}
\providecommand{\abs}[1]{\left\lvert#1\right\rvert}

\makeindex

\newcommand{\iek}[1]{\left(#1\right)}

\newcommand{\kiek}[1]{\left[#1\right]}
\providecommand{\vect}[1]{\pmb{#1}}

\begin{document}

\title{Rotating hematite cube chains}

\author{M.\ Brics}
\author{V.\ Šints}
\author{G.\ Kitenbergs}
\author{A.\ Cēbers}
\affiliation{%
 MMML lab, Department of Physics,
University of Latvia, Jelgavas 3, Rīga, LV-1004, Latvia
}

\date{\today}

\begin{abstract}

Recently a two-dimensional chiral fluid was experimentally demonstrated. It was obtained from cubic-shaped hematite colloidal particles placed in a rotating magnetic field. Here we look at building blocks of that fluid, by analyzing short hematite chain behavior in a rotating magnetic field.  We find equilibrium structures of chains in static magnetic fields and observe chain dynamics in rotating magnetic fields. We find and experimentally verify that there are three planar motion regimes and one where the cube chain goes out of the plane of the rotating magnetic field. In this regime we observe interesting dynamics --- the chain rotates slower than the rotating magnetic field. In order to catch up with the magnetic field, it rolls on an edge and through rotation in the third dimension  catches up with the magnetic field. The same dynamics is also observable  for a single cube when gravitational effects are explicitly taken into account. 
\end{abstract}


\maketitle


\section{Introduction}
\label{sec:intro}

\begin{figure}[ht]
\includegraphics[width=0.48\columnwidth]{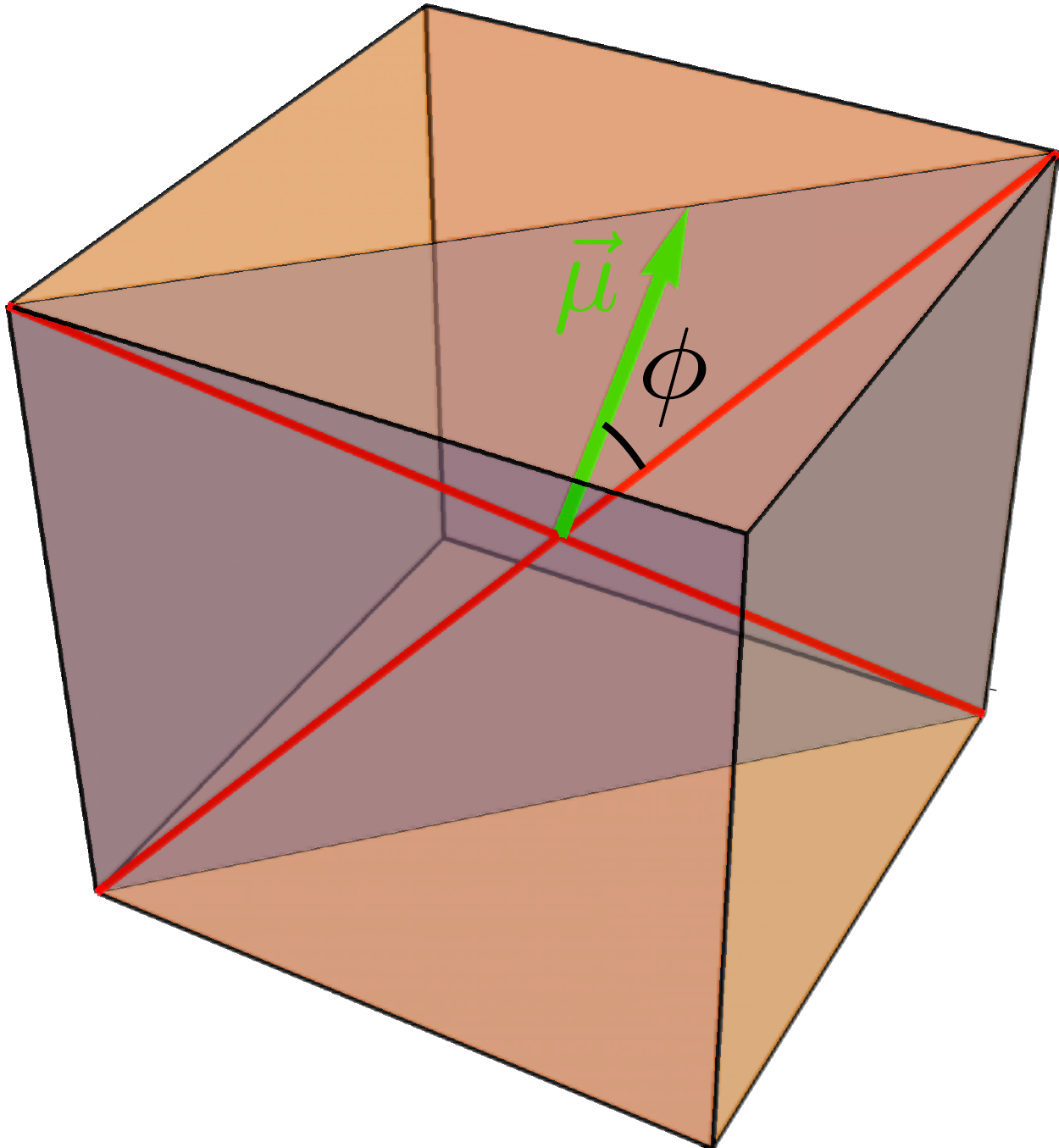}
\caption{Orientation of magnetic moment in a hematite cube. The angle $\phi=12^\circ$ is in the plane defined by two diagonals  and the magnetic moment $\vect \mu$ points to the face.}
\label{fig:cube}
\end{figure}

At room temperature hematite is a weak ferromagnetic material with an unorthodox magnetization orientation: for cubic-shaped hematite particles the magnetic moment with a cube's diagonal  makes an angle 12$^\circ$ (see Fig.~\ref{fig:cube}) in the plane defined by two diagonals \cite{Philipse2018, Brics2022}. Thus, leading to interesting physical effects. In static magnetic field and low concentrations cubic shaped hematite particles arrange in kinked chains \cite{Philipse2018, Brics2022}. If concentration is increased \cite{Rossi_phd, Petrichenko_2020}, swarms are formed. If we let a lot of swarms interact, we can observe  a two-dimensional chiral fluid in a rotating magnetic field \cite{Soni}. The chiral fluid consists of individual cubes and short hematite chains (usually two and three cube chains) which are interacting in a rotating magnetic field \cite{Petrichenko_2020}.

In the scientific literature there are also other interesting experiments with hematite colloids formed by cubic particles. In the article \cite{Chen2021} authors investigated application of cube-shaped hematite microrobots for microblocks and impurities sweep in blood vessels. There approximately $2\,\mathrm{\mu m}$ large cube-shaped hematite particles were guided by the rotating magnetic field through introduced rolling motion. They  showed that hematite cubes can overcome obstacles and push small objects.  Motile structures formed by microrollers which were created by micron sized polymer colloids with embedded hematite cubes were demonstrated in \cite{Driscoll_2016}. In \cite{Aubret_2018} authors demonstrated targeted assembly and synchronization of self-spinning microgears or rotors made of hematite cubes and chemically inert polymer beads. In \cite{Castillo_2014} was investigated a potential application of hematite colloidal cubes for the enhanced degradation of organic dyes. In \cite{Palacci_2013} were examined the formations of light activated two-dimensional “living crystals”.

In this article we investigate the building blocks of chiral fluid demonstrated in \cite{Soni}, i.e. individual cube and short hematite chain behavior in a rotating magnetic field. We perform analytical calculations and simulations which we later confirm with experiments. To determine how important gravity effects are two models were developed, one with explicit gravity treatment and one without. 

The content of the paper is divided into five sections. The Sec.~\ref{sec:intro}  is introduction followed by Sec.~\ref{sec:theor} where  theoretical methods are described. The theoretical and experimental results are given in Sec.\ref{sec:theor_res} and Sec.\ref{sec:exp} respectively and conclusions
in Sec.~\ref{sec:concl}. This article also contains supplementary material --- videos to better illustrate different motion modes.  

\section{Theoretical methods}
\label{sec:theor}

\begin{figure}[htbp]
\includegraphics[width=0.5\columnwidth]{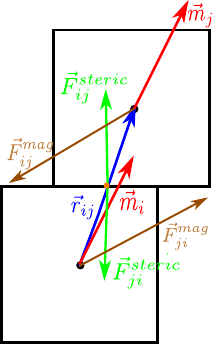}
\caption{Forces acting on a two-cube chain.}
\label{fig:force}
\end{figure}

To theoretically describe the behaviour of hematite chains in a rotating magnetic field we use a microscopic model where motion of each hematite particle is described. The equations of motion (EOMs) are derived using the Newton mechanics approach. Only the essential forces are introduced to keep the number of parameters minimal and ease interpretation. For each hematite cube  in an external homogeneous magnetic field $\vec{B}$ with magnetic moment $\vec{m}$, mass $\mathtt{m}$, moment of inertia tensor $\vect{I}$, the force and torque balance is considered. From the force balance (two-particle forces are shown in Fig.~\ref{fig:force})  we obtain that 
\begin{equation}
 {\mathtt{m}_j\frac{\mathrm{d}}{\mathrm{d} t}\vec{v}_j}=\vec{F}^{HD}_{j}+
      \sum_i\iek{\vec{F}^{mag}_{ij}+\vec{F}^{steric}_{ij}},
      \label{eq:forces}
\end{equation}
 where $\vec{F}^{HD}_{j}$ is the hydrodynamics force acting on the particle $j$, $\vec{F}^{mag}_{ij}$ is the magnetic force produced by particle $i$, and  $\vec{F}^{steric}_{ij}$ is the reaction force that ensure that particles do not overlap. The corresponding equation obtained from torque $T$ balance then reads:
 \begin{equation}
 {\vect{I}_j\frac{\mathrm{d}}{\mathrm{d} t}\vec{\Omega}_j}=\vec{m}_j\times{\vec{B}}+ \vec{T}^{HD}_{j}+
      \sum_i\iek{\vec{T}^{mag}_{ij}+\vec{T}^{steric}_{ij}},
      \label{eq:torques}
\end{equation}
 where $\vec{m}_j\times{\vec{B}}$ is magnetic torque produced by the external field and $\vec{T}^{HD}$, $\vec{T}^{mag}$, $\vec{T}^{steric}$ are corresponding torques of forces in  Eq.~\ref{eq:forces}. Later, in order to explicitly incorporate gravity effects in the model, we add the buoyancy force $\vec{F}^b_j=(\rho_h-\rho_s)g a_0^3$, the reaction force $\vec{F}^{wall}_{j}$ and torque $\vec{T}^{wall}_{j}$ with the bottom of capillary acting on each particle. Here $g=9.81\,\mathrm{m/s^2}$ is the gravitational acceleration, $a_0\approx1.5\,\mathrm{\mu m}$ is the edge length of a hematite cube and $\rho_h=5.25\,\mathrm{ g/cm^3}$ and $\rho_s=1.00\,\mathrm{ g/cm^3}$ are densities of hematite and solvent, which in our case is water.  However, in this work we do not consider any other forces like friction between cubes and between a cube and capillary or thermal fluctuations.
 
Note that, in the general case, the knowledge of $\vect{\Omega}$ at a specific time point is not sufficient to determine the orientation of the particle. Thus, for non-spherical objects, we lack information to calculate reaction forces and torques that ensure that particles do not overlap. To overcome this issue rotational matrices or quaternions at each time step have to be calculated  \cite{Omelyan1998,Kou2018}. Here we choose to use quaternions $\vect{q}_i$ \cite{quat} as this approach requires to calculate time evolution of less dimensional quantity and equation of motion
\begin{equation}
 \frac{\mathrm d}{\mathrm{d} t} \vect{q}_i=\frac{1}{2}\begin{pmatrix}
0 & -\Omega_z^i&  \Omega_y^i &  \Omega_x^i \\
\Omega_z^i & 0 & -\Omega_x^i & \Omega_y^i \\
-\Omega_y^i  & \Omega_x^i & 0 & \Omega_z^i  \\
-\Omega_x^i & -\Omega_y^i & -\Omega_z^i & 0 
\end{pmatrix} \vect{q}_i =\vect{Q}(\vect{\Omega}_i)\vect{q}_i
\label{eq:quatx}
\end{equation}
is always stable \cite{Omelyan1998,Kou2018}, unlike EOM for Euler angles. 
{ The quaternion is a four dimensional quantity, which satisfies the normalization condition and provides a convenient representation of spatial orientations and rotations of elements in three dimensional space (corresponds to a rotation matrix). E.g., rotation around the axis $\vect{u}=(u_x, u_y, u_z)$ by an angle $\theta$ can be expressed with quaternion $\vect{q}=(u_x\sin\frac{\theta}{2},u_y\sin\frac{\theta}{2},u_z\sin\frac{\theta}{2}, \cos\frac{\theta}{2})$.}
Note that the angular velocities in the laboratory frame are used. Quaternions are implemented using  scalar last notation as internally stored in a C++ template library for linear algebra Eigen \cite{eigen}.

For the experimental conditions \cite{Soni, Petrichenko_2020} we are interested in, it turns out that corresponding Reynolds numbers $Re<<1$ and inertial terms are negligible (${\mathtt{m}\frac{\mathrm{d}}{\mathrm{d} t}\vec{v}}<<\vec{F}^{HD}$). Thus, for our calculation we neglect inertial terms and use the Stokes approximation. For hydrodynamics forces and torques, to keep equations analytically analyzable,  we use linear velocity drag approximation which for a cubic shaped particle reads \cite{Okada2018}: 
\begin{align}
 {\vec{F}^{HD}_{j}}&=-\xi \vect{v}_{j}; \quad \xi\approx3 \pi \eta a_0 \cdot 1.384;\\
 {\vec{T}^{HD}_{j}}&=-\zeta \vect{\Omega}_j ;\quad  \zeta= \pi \eta a^3_0 \cdot 2.552,
\end{align}
where $\eta=1.0\,\mathrm{mPa\cdot s}$ is the viscosity of the solvent, which in our case is water, and $\xi$ and $\zeta$ are drag and rotational drag coefficients respectively. 

To calculate { particle magnetic} interactions as in \cite{Brics2022} we use the dipole approximation {color{blue} since} qualitatively the results are the same \cite{Brics2022}, despite the fact that quantitative differences in specific arrangements are up to 18\%.  The magnetic-magnetic particle interaction expressions for the force and torque reads: 
\begin{equation}
\vec{F}^{mag}_{ij}=\frac{3\mu_0m^2}{4\pi r_{ij}^4}\vect{\tilde{F}}_{ij}^{mag}; \quad \vec{T}^{mag}_{ij}=\frac{3\mu_0m^2}{4\pi r_{ij}^3}\vect{\tilde{T}}_{ij}^{mag};  
\end{equation}
\begin{equation}
\begin{split}
 \vect{\tilde{F}}_{ij}^{mag}=&\vect{\hat r}_{ij}(\vect{\hat m}_i\cdot \vect{\hat m}_j) 
 +
  \vect{\hat m}_i (\vect{\hat r}_{ij}\cdot \vect{\hat m}_j)
   \\&+ 
  \vect{\hat m}_j (\vect{r}_{ij}\cdot \vect{\hat m}_i)-5\vect{\hat r}_{ij}(\vect{\hat r}_{ij}\cdot \vect{\hat m}_i)(\vect{\hat r}_{ij}\cdot \vect{\hat m}_j),
\end{split} 
\end{equation}

\begin{equation}
  \vect{\tilde{T}}_{ij}^{mag}=(\vect{\hat r}_{ij}\cdot \vect{\hat m}_i)(\vect{\hat m}_j \times\vect{\hat r}_{ij})+\frac{1}{3}(\vect{ \hat m}_i\times \vect{\hat m}_j),
\end{equation}
where $r_{ij}$ is the vector between $i$-th and $j$-th particle centers (shown in Fig.~\ref{fig:force}), $\mu_0=4\pi \cdot 10^{-7}\,\mathrm{H/m}$ is the { magnetic permeability of vacuum and quantities with $\tilde{}$ denote dimensionless variables apart from unit vectors, which are denoted with  $\hat{}$,  e.g. $\vect{r}_{ij}=r_{ij}\hat{\vect{r}}_{ij}$.}   

\begin{figure}[ht]
\includegraphics[width=0.5\columnwidth]{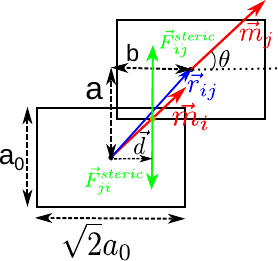}
\caption{Reaction forces between two cubes. The schematic view of the two-particle chain from above. As a cube lies on an edge, the projection to the base is  a rectangle with sides $a_0$ and $\sqrt{2}a$, if the tilt angle (the angle between the base and cube's face) is  45$^\circ$.}
\label{fig:steric}
\end{figure}

Already mentioned reaction forces $\vec{F}^{steric}_{ij}$ and corresponding torques $\vec{T}^{steric}_{ij}$ are added to avoid cube overlap. The results do not depend on the choice of exact expression for reaction forces whenever the model for reaction forces is reasonably chosen. Therefore, here we use some power function which has only a repulsive part:
\begin{align}
\vec{F}^{steric}_{ij}&=\frac{3\mu_0m^2}{4\pi a_{0}^4}\vect{\tilde{F}}_{ij}^{steric}; \quad
\vec{T}^{steric}_{ij}=\frac{3\mu_0m^2}{4\pi a_{0}^3}\vect{\tilde{T}}_{ij}^{steric};\\
\vect{\tilde{F}}_{ij}^{steric}&=\frac{A}{a_0^2}\kiek{\iek{\frac{1}{1-4\theta_H(\frac{a}{a_0}-1)}}^{13}-1}\hat{\vect{F}}^{steric}_{ij}; \\
\vect{\tilde{T}}_{ij}^{mag}&= \frac{\vec{d}}{a_0}\times \vect{\tilde{F}}_{ij}^{steric},
\end{align}
where $\theta_H$ is Heaviside step function, $A$ is the area with which cubes touch and quantities $a$ and $\vec{d}$ are defined in Fig.~\ref{fig:steric}. This approach to calculate reaction forces and torques becomes, however, computationally very demanding for non-planar motion. Especially, if one takes into account that particles used in  experiments  are with rounded corners, i.e. superballs \cite{Brics2022, Rossi_phd, Philipse2018}. Thus, to ease the computational task we reconstruct superballs out of spheres and calculate reaction forces as in \cite{Kantorovich2,Brics2022}. Each cube we replace with 93 spheres as in \cite{Brics2022} and calculate repulsion forces for every sphere with { every sphere of other cube's}. For steric repulsion we are using Weeks-Chandler-Anderson potential \cite{WCA}. Also in a similar way we calculate reaction forces with the bottom of capillary for the model with gravity.

Combining all expressions the EOM for dimensionless variables reads:
\begin{align}
\begin{split}
 \vect{\tilde v}_j&=ks\sum_{i} \left(\frac{1}{\tilde{r}_{ij}^4}\vect{\tilde{F}}_{ij}^{mag}+\vect{\tilde{F}}_{ij}^{steric}\right),\\
   \vect{\tilde \Omega}_j&=\vect{\hat{m}}_j\times \vect{\hat B}+s\sum_{i} \left(\frac{1}{\tilde{r}_{ij}^3}\vect{\tilde{T}}_{ij}^{mag}+\vect{\tilde{T}}_{ij}^{steric}\right),\\
   \frac{\mathrm d}{\mathrm{d} \tilde t} \vect{q}_j&=\vect{Q}(\vect{\tilde \Omega}_i)\vect{q}_j, \label{eq:quat}
\end{split}
\end{align}
where nondimensionalization for time $\tilde{t}=\frac{\zeta t}{mB}$ and distance $\tilde{\vect{r}_{ij}}=\frac{ \vect{r}_{ij}}{a_0}$  is used leading to dimensionless variables $\tilde{\vect{v}}=\frac{\vect{v}\zeta} { mB a_0}$  and $\tilde{\vect{\Omega}}=\frac{\vect{\Omega} \zeta }{mB}$. The EOM has two controlparameters $s$ and $k$, from which only 
\begin{equation}
 s=\frac{3\mu_0m }{4\pi a_0^3 B }\approx 6.6\frac{B_c}{B}; \quad B_c\approx0.1\,\mathrm{mT}
\end{equation}
is adjustable in experiments by changing the magnitude of the external magnetic field. The parameter
\begin{equation}
  k=\frac{ \zeta}{a_0^2 \xi}\approx0.614
\end{equation}
is drag coefficient ratio and thus is fixed in experiments .

In the case of gravity we have one additional parameter
\begin{equation}
G_m=\frac{4 \pi(\rho_h-\rho_s)g a_0^7}{3\mu_0m^2},    
\end{equation}
which is the ratio of buoyant forces to magnetic forces. The EOMs in this case reads:
\begin{align}
\begin{split}\vect{\tilde v}_j=& ks\sum_{i} \left(\frac{1}{\tilde{r}_{ij}^4}\vect{\tilde{F}}_{ij}^{mag}+\vect{\tilde{F}}_{ij}^{steric}\right)\\
&+ks\iek{G_m\hat{\vect{g}}+\vect{\tilde{F}}_{j}^{wall}},\\
\vect{\tilde \Omega}_j=&\vect{\hat{m}}_j\times \vect{\hat B}+s\sum_{i} \left(\frac{1}{\tilde{r}_{ij}^3}\vect{\tilde{T}}_{ij}^{mag}+\vect{\tilde{T}}_{ij}^{steric}\right)\\
   &+s\vect{\tilde{T}}_{j}^{wall},\\
   \frac{\mathrm d}{\mathrm{d} \tilde t} \vect{q}_j=&\vect{Q}(\vect{\tilde \Omega}_i)\vect{q}_j. \label{eq:quat1}
\end{split}   
\end{align}

\section{Theoretical results}
\label{sec:theor_res}
\subsection{Single particle}

In the case of a single cube (all other cubes are sufficiently far away) and no gravity effects, equations read: 
\begin{align}
   \vect{\tilde v}&=0\,,\\
   \vect{\tilde \Omega}&=\vect{\hat{m}}\times \vect{\hat B}.
  \end{align}
There is no dependence on control parameters $s$ and $k$ as well as on the shape of the particle.

\begin{figure}[ht]
\includegraphics[width=0.8\columnwidth]{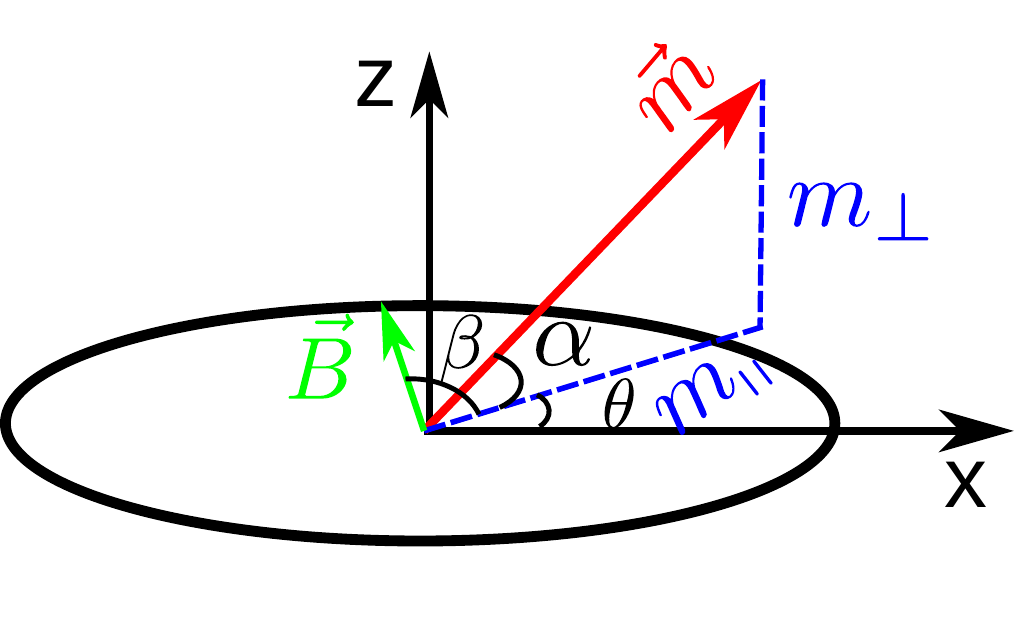}
\caption{The magnetic moment in rotating magnetic field.}
\label{fig:mom_rot}
\end{figure}

If we  we assume that  magnetic field rotates in xy-plane, i.e. $\vect{\hat B}=\cos(\tilde \omega \tilde t)\vect{e}_x+\sin(\tilde \omega \tilde t)\vect{e}_y$ with $\tilde{\omega}=\frac{{\omega} \zeta }{ mB}$, it is beneficial to introduce two angles $\theta$ and $\alpha$ (see Fig.~\ref{fig:mom_rot}) to describe the motion of the cube . $\alpha$ is the angle the
magnetic moment makes with the plane of external magnetic field and $\theta$ is the angle which the magnetic moment's projection in the plane of the rotating magnetic field makes with $x$-axis. In this case $\vect{\hat m}=\cos(\theta)\cos(\alpha) \vect{e}_x+\sin(\theta)\cos(\alpha)\vect{e}_y+\sin(\alpha)\vect{e}_z$ and  the angular velocity can be expressed as $\vect{\tilde \Omega} = \vect{\hat m} \times \dot{\hat{\vect{m}}}$.
The EOMs, which previously were derived in \cite{Palkar2019}, for a single cube in this case reads 
\begin{align}
\dot \alpha&=\cos(\tilde \omega \tilde t-\theta)\sin(\alpha)\,,\\
 \dot \theta&=\sin(\tilde \omega \tilde t-\theta)/\cos(\alpha)\,.
\end{align}
To analyze this equation it is beneficial to introduce the lag angle $\beta=\tilde \omega \tilde t-\theta$ as for variable $\beta$ unlike for $\theta$ there are fixed points.  The OEM for the lag angle reads
\begin{align}
 \dot \alpha&=-\sin(\alpha)\cos(\beta)\,,\label{eq:lag1}\\ 
 \dot \beta&=\tilde \omega-\sin(\beta)/\cos(\alpha)\,.
 \label{eq:lag}
\end{align}

 The Eqs.~\ref{eq:lag1} ~- \ref{eq:lag} has four stationary points $P(\alpha, \beta)$:  $P_1=\{0,$ $\mathrm{asin}(\tilde \omega)\}$, $P_2=\{0$, $\pi-\mathrm{asin}(\tilde \omega)\}$ and $P_3=\{\mathrm{acos}(1/\tilde \omega)$, $\frac{\pi}{2} \}$, $P_4=\{-\mathrm{acos}(1/\tilde \omega)$, $\frac{\pi}{2} \}$. The first two fixed points exist if $\abs{\tilde \omega} \leq \tilde{\omega}_c$ and the last two  when  $\abs{\tilde \omega} \geq \tilde{\omega}_c$, where the critical frequency $\tilde{\omega}_c=1$ .
 
\begin{figure}[ht]
\includegraphics[width=0.8\columnwidth]{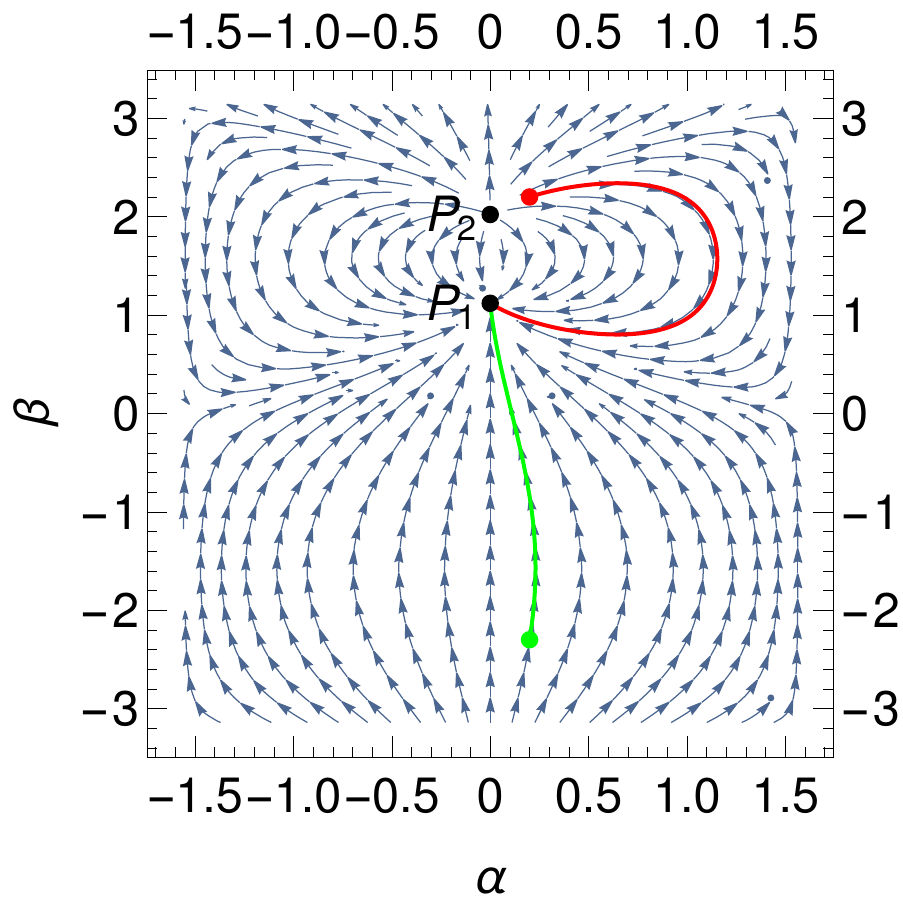}\\
\includegraphics[width=0.95\columnwidth]{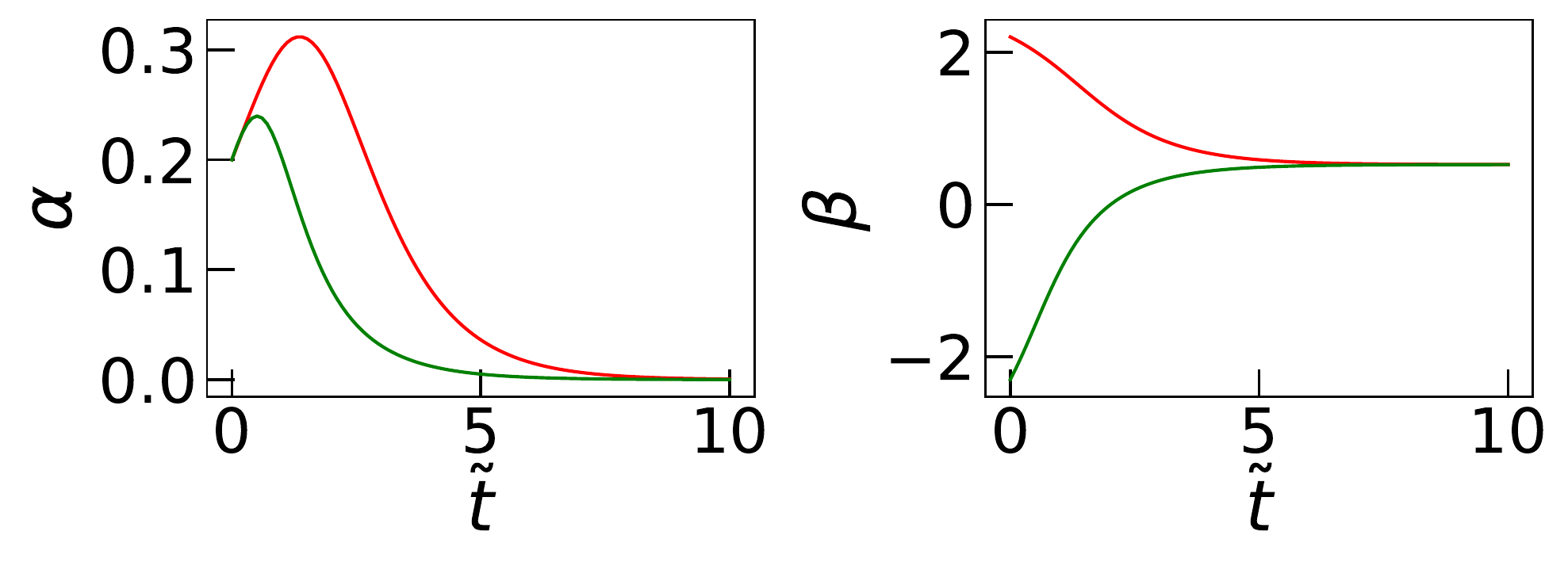}
\caption{A phase portrait of single magnetic cube in rotating magnetic field  calculated using Eqs.~\ref{eq:lag1} ~- \ref{eq:lag} with $\tilde \omega = 0.5$. The time evolution of two particular trajectories (red and green) is plotted  below. Gravity effects are not taken into account.}
\label{fig:PhasePlot1}
\end{figure}

If $\abs{\tilde \omega} < 1$ then there are two stationary  points $P_1$ and $P_2$. As one can see from the Fig.~\ref{fig:PhasePlot1}, the point $P_1$ is stable and $P_2$ unstable, thus acting as a think and source. Independent of initial conditions after some transition time stationary point $P_1$ is reached. Any perturbation is suppressed. The cube rotates synchronously with the frequency of the external magnetic field and the magnetic moment is in the plane of the rotating magnetic field. It lags the direction of the magnetic field by an angle $\beta_j=\mathrm{asin}(\tilde \omega)$. By increasing frequency the points $P_1$ and $P_2$ move closer to each other and at $\abs{\tilde \omega}=1$ coincide and for $\abs{\tilde \omega} > 1$ disappear. 

\begin{figure}[ht]
\includegraphics[width=0.8\columnwidth]{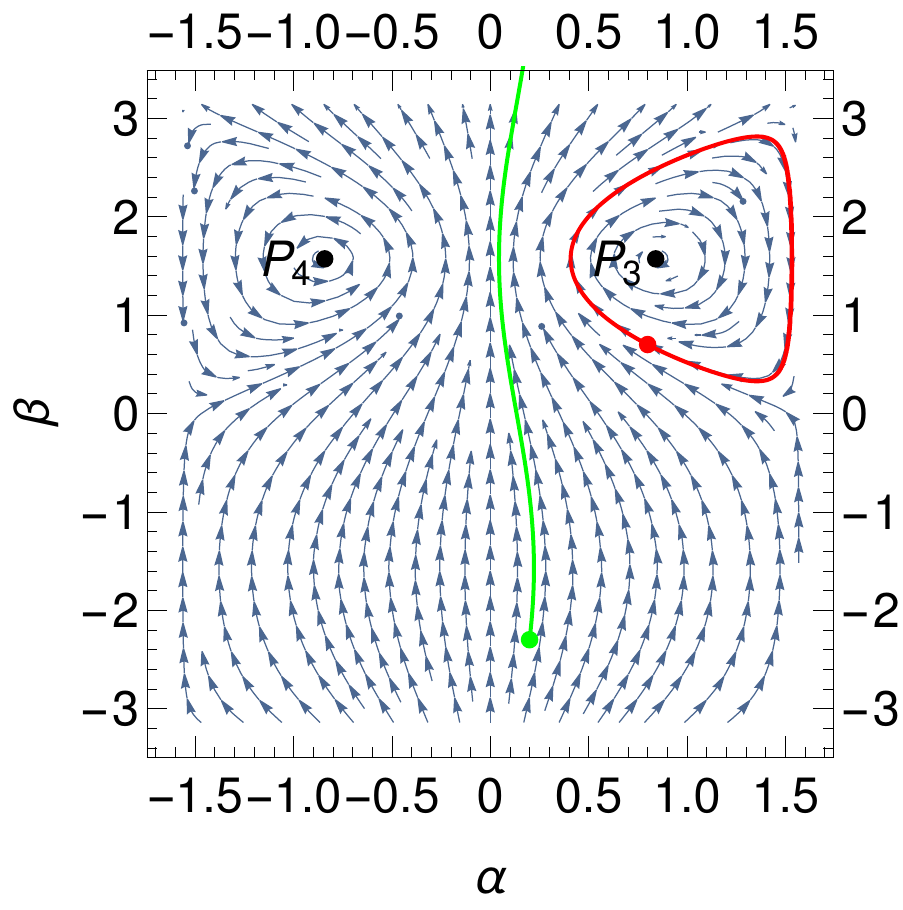}\\
\includegraphics[width=0.95\columnwidth]{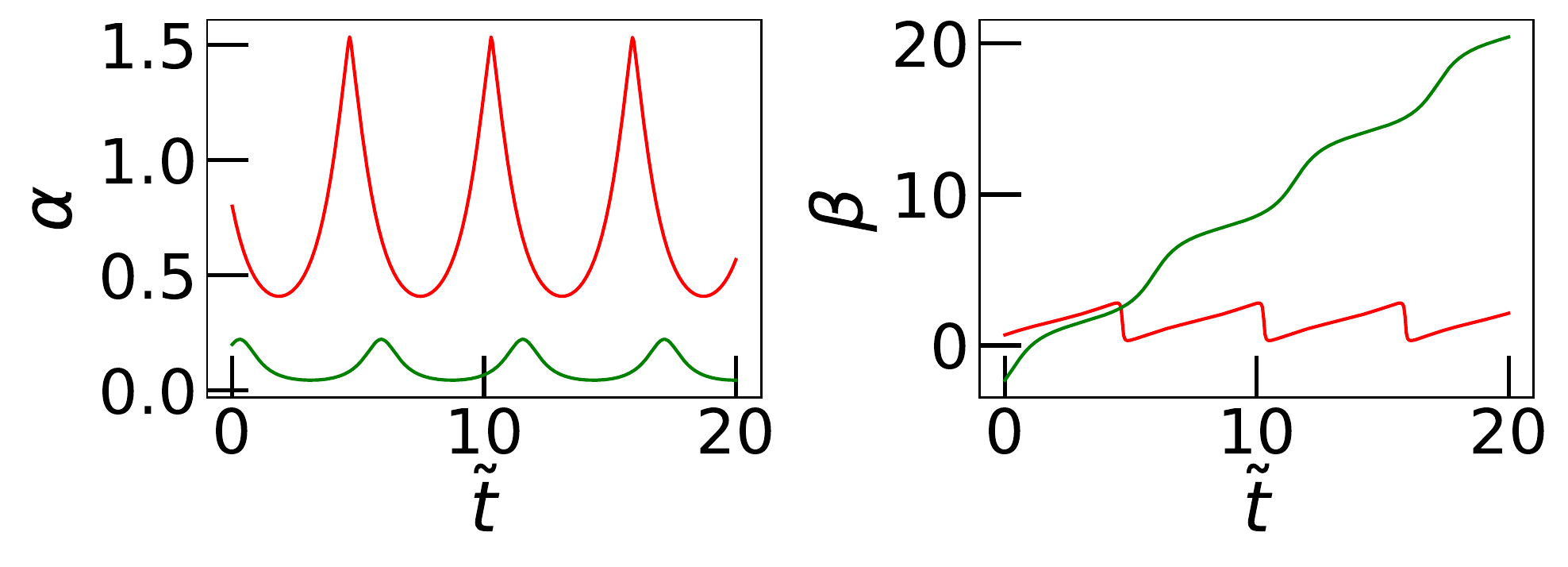}
\caption{A phase portrait of single magnetic cube in rotating magnetic field  calculated using Eqs.~\ref{eq:lag1} ~- \ref{eq:lag} with $\tilde \omega = 1.5$. The time evolution of two particular trajectories (red and green) is plotted  below. Gravity effects are not taken into account.}
\label{fig:PhasePlot2}
\end{figure}

Situation gets more interesting when $\abs{\tilde \omega} > 1$. In this case there are also two stationary points $P_3$ and $P_4$. However, both points  $P_3$ and $P_4$ are neutrally stable and act as centers for rotation, as one can see from the Fig.~\ref{fig:PhasePlot2}. No stationary solution is possible except points $P_3$ and $P_4$, and the system can reach $P_3$ and $P_4$ only if it initially was there. Depending on the initial condition, two scenarios are possible. One option is that both angles periodically oscillate. The other option is that there is rotation around one of the stationary points. This means that in the first case  the motion is not anymore synchronous with the external magnetic field and we observe back-and-forth motion where the magnetic moment can be out of the plane of rotation. In the second case  we observe precession of the magnetic moment. The motion is asynchronous except for stationary points  $P_3$ and $P_4$. { The motion modes: synchronous rotation, precession of the magnetic moment, and back-and forth rotation arranged from left to right are shown in Video1 \cite{Video0} in rotating frame, which is rotating with the magnetic field (top row) and laboratory frame (bottom row). Synchronous rotation and precession is easier identified in the rotating frame but back-and-forth rotation in the laboratory frame. Note that cube can be rotated by an arbitrary angle around an axis parallel to the magnetic field.}

\begin{figure}[ht]
\includegraphics[width=0.9\columnwidth]{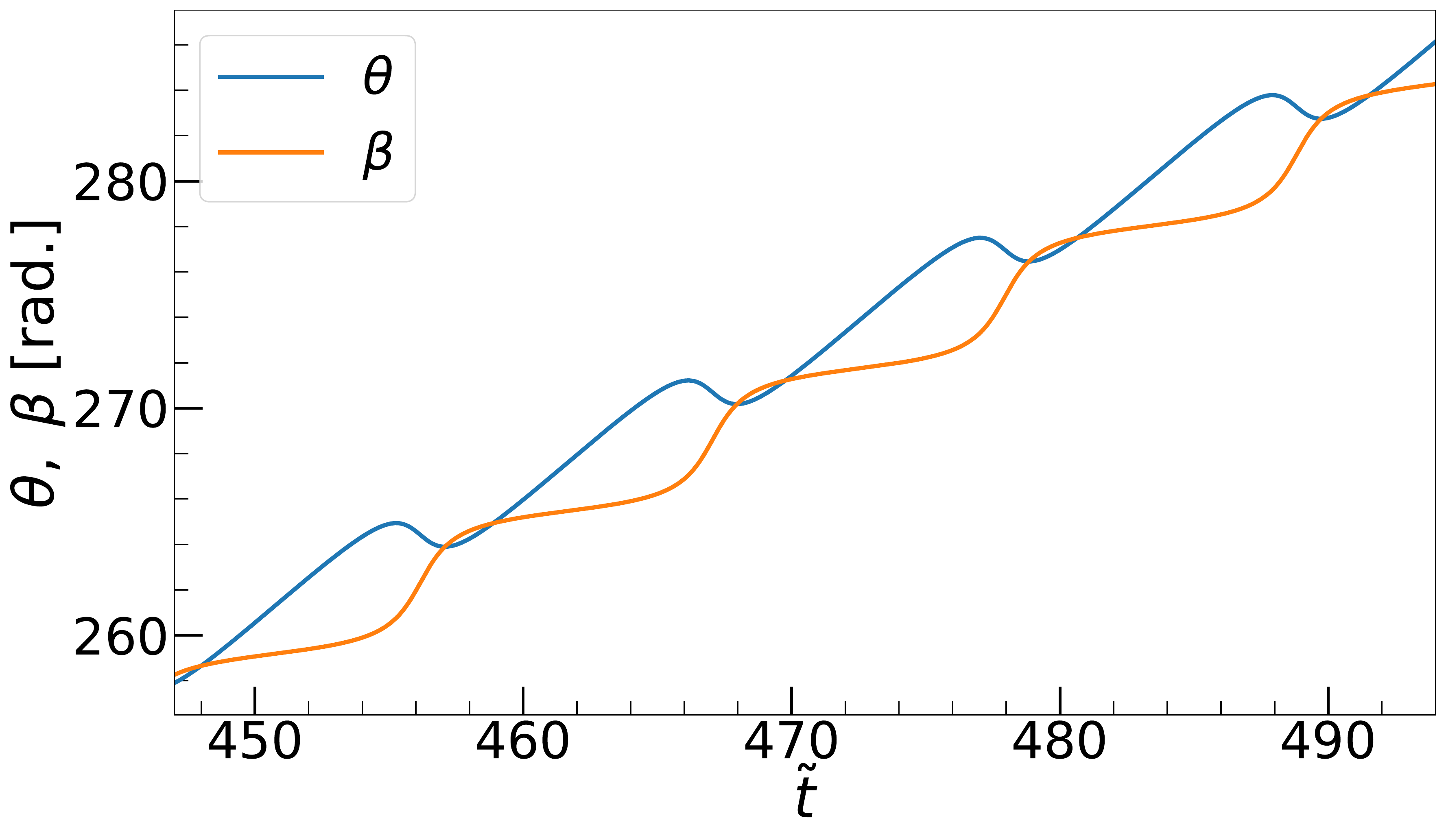}
\caption{The time evolution after transition period of orientation angle $\theta$ and lag angle $\beta$ calculated using Eqs.~\ref{eq:lag1} ~- \ref{eq:lag} with $\tilde \omega=2/\sqrt{3}$, which correspond to $W_n=1$.}
\label{fig:wn_1}
\end{figure}

From Fig. \ref{fig:PhasePlot2} follows that, if the magnetic moment is initially in the plane of the rotating magnetic field, it will always remain there. As this situation  is analytically solvable we examine it in more detail. 

When one limits $\beta_j\in[-\pi, \pi]$ then
 \begin{equation}
  \beta_j=2 \mathrm{atan}\left[\frac{1+\sqrt{\tilde{\omega}^2-1}\tan(\sqrt{\tilde{\omega}^2-1}\frac{\tilde t-\tilde t_0}{2}) }{\tilde \omega} \right]\,, 
 \end{equation}
with period $\tilde T=\frac{2\pi}{\sqrt{\tilde \omega^2-1}}$.                                                                                                                         
The cube oscillates forth and back and during one period makes
\begin{equation}
 W_n=\frac{1}{2\pi}\int_0^{\tilde T} \mathrm{d}\, \theta_j=\frac{1}{2\pi}\int_0^{2\pi}\frac{\dot \theta_j}{\dot \beta_j} \mathrm{d}\, \beta_j=\frac{\abs{\tilde{\omega}}}{\sqrt{\tilde \omega^2-1}} -1
\end{equation}
winds around its axis of rotation which is perpendicular to the plane of the rotating magnetic field. The particular trajectory for the rotational field frequency $\tilde \omega=2/\sqrt{3}$, which corresponds to $W_n=1$, can be seen in Fig.~\ref{fig:wn_1}. When $W_n>1$, then lag angle $\beta$ increases faster than orientation angle $\theta$. The opposite is observed if $W_n<1$. 

\subsubsection{Gravitation}

\begin{figure}[ht]
\includegraphics[width=0.9\columnwidth]{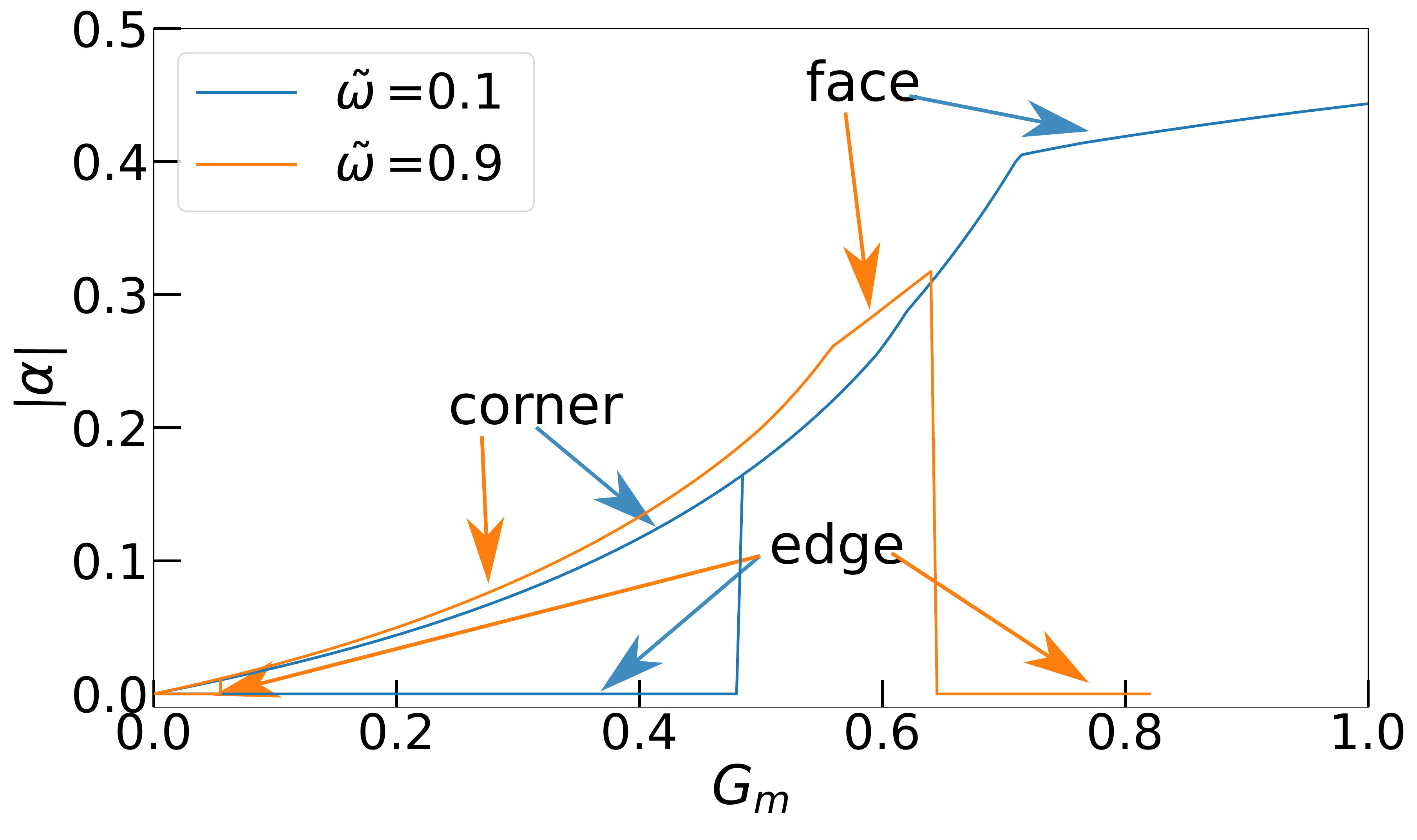}
\caption{Stationary solutions for angle $\alpha$ at $\tilde\omega=0.1$ and $\tilde\omega=0.9$ for increasing gravity parameter $G_m$, obtained from long-time solutions of Eq.~\ref{eq:quat1} with $k=0.614$ and $s=0.94$ and for fixed gravity parameter $G_m$. In synchronous rotation cube with rounded corners can rotate either on a corner, face, or an edge.}
\label{fig:gs1}
\end{figure}

\begin{figure}[ht]
\includegraphics[width=0.9\columnwidth]{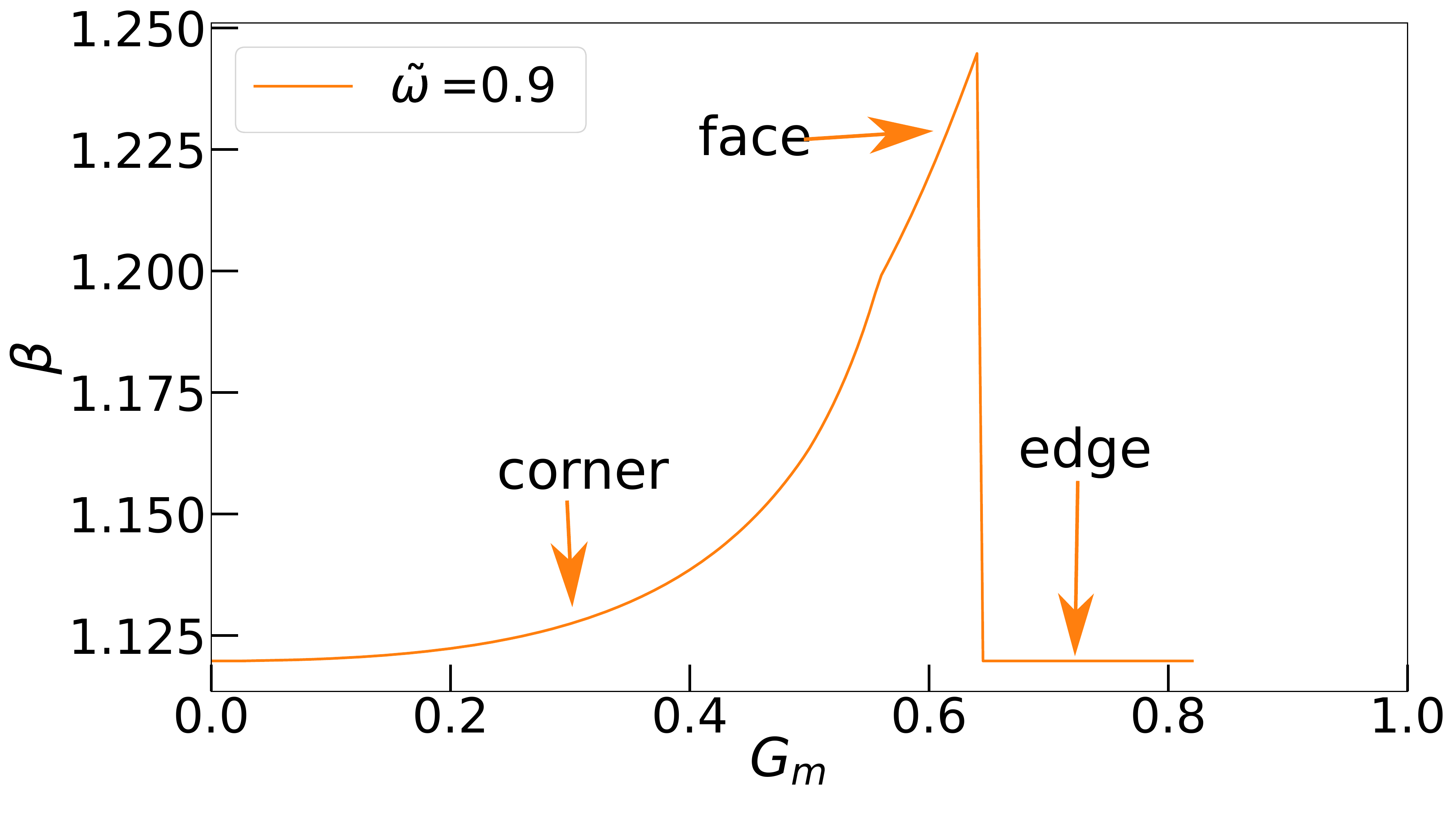}
\caption{Stationary solutions for lag $\beta$ at $\tilde\omega=0.9$ for increasing gravity parameter $G_m$ for the same parameter values as Fig.~\ref{fig:gs1}.}
\label{fig:gs2}
\end{figure} 

Due to an unorthodox orientation of magnetic moment in a hematite cube, if the magnetic moment is in the plane of the rotating magnetic field, cube lies on an edge \cite{Brics2022}. Thus, it is not the minimum of potential gravitational energy. Therefore, if maximal gravitational torque $T_g^{max}=\frac{(\rho_h-\rho_s)ga^4}{2}$ is larger than maximal magnetic torque $T_m^{max}=MBa^3$,  the cube lies on a face and the magnetic moment is always  out of the plane of the rotating magnetic field. For hematite cube with $a\approx 1.5 \,\mathrm{\mu m}$, density $\rho_h=5.25\,\mathrm{ g/cm^3}$,  solvent density  $\rho_s=1.00\,\mathrm{ g/cm^3}$ , and permanent magnetization $M=2.2\times10^3\,\mathrm{A/m}$ ($m=Ma^3$) one finds that $T_g^{max}>T_m^{max}$ if $B<15 \,\mathrm{\mu T}$. This is much smaller magnetic field than those used both here and in earlier experiments \cite{Petrichenko_2020, Soni}. We use $B_{exp}\in [0.3; 3]\,\mathrm{ mT}$. However, as gravitational effects may change dynamics, especially of neutrally stable points $P_3$ and $P_4$, we examine them in detail.

To do so we add to our model, the buoyancy  force $\vect{F}^b$ and the reaction force $\vect{F}^{wall}$ and torques $\vect{T}^{wall}$ with the bottom of capillary and solve Eq.~\ref{eq:quat1} with fixed values $k=0.614$ and $s=0.94$. Note, that even for a single cube to calculate reaction forces and torques we solve the  equation for quaternion Eq.~\ref{eq:quat}, thus we are evolving in time a seven dimensional object (3 coordinates and 4 components of quaternion). Although, only five of them change in time as ($x$ and $y$) remain fixed as we do not have friction nor vertical walls.

For the shape of hematite cubes used in experiments (superballs --- cubes with rounded corners) one finds that dynamics gets more complicated. From our simulations we see that an individual cube with rounded corners rotates on an edge, face, corner or undergoes a complicated 3D motion (will be discussed later in this section). The critical value of $\tilde \omega$, which separates the synchronous motion of cube from asynchronous,  changes  to $\tilde \omega_c=|\cos(\alpha)|$ in the case when a cube rotates on the face and magnetic torque can not overcome gravitational one. Note that here we neglect stationary points $P_3$ and $P_4$ as starting from random initial conditions the probability to reach them is zero.

For $\tilde \omega<\tilde \omega_c$ there are three possibilities: either a cube rotates on the edge, face or corner {(see Video2 \cite{Video0})}. The magnetic moment is in the plane of the rotating field only if the cube rotates on the edge. The stationary solutions for angles $\alpha$ and $\beta$ with increasing gravity parameter $G_m$ are shown in Figs. ~\ref{fig:gs1}, \ref{fig:gs2}. From Fig.~\ref{fig:gs1} one finds that for 
weak gravity there are two stationary solutions of synchronous rotation. Cube as before can rotate on an edge with magnetic moment in the plane of rotations or on a corner with magnetic moment pointing out of the plane of rotating magnetic field. Whether a cube rotates one a face or edge depends on the initial conditions and $\tilde\omega$.  At some point the rotation on the edge becomes unstable and the cube starts to rotate on the corner. This happens faster the larger  is the rotational frequency. The momentum goes out of the plane of the rotating magnetic field. The angle $\alpha$ for fixed gravity parameter depends on the shape of the cube (how much corners are rounded) and rotation frequency $\tilde \omega$. If the gravity parameter is increased the cube starts to rotate on the face. For $\tilde \omega>0.81$ there is again a region where rotation on the edge becomes the only stable solution. The reason is that in this interval, the magnetic moment can not anymore balance the gravitational moment when the cube rotates on a face, but can when the cube rotates on an edge.  This is the case because the magnetic torque is smaller when the cube rotates on a face. For $\tilde \omega=0.9$ and $G_m>0.83$ synchronous motion is not anymore observed. There the motion, which is a blend of back-and-forth motion and precession is observed. This motion is described in the next paragraph.

\begin{figure}[ht]
\includegraphics[width=0.99\columnwidth]{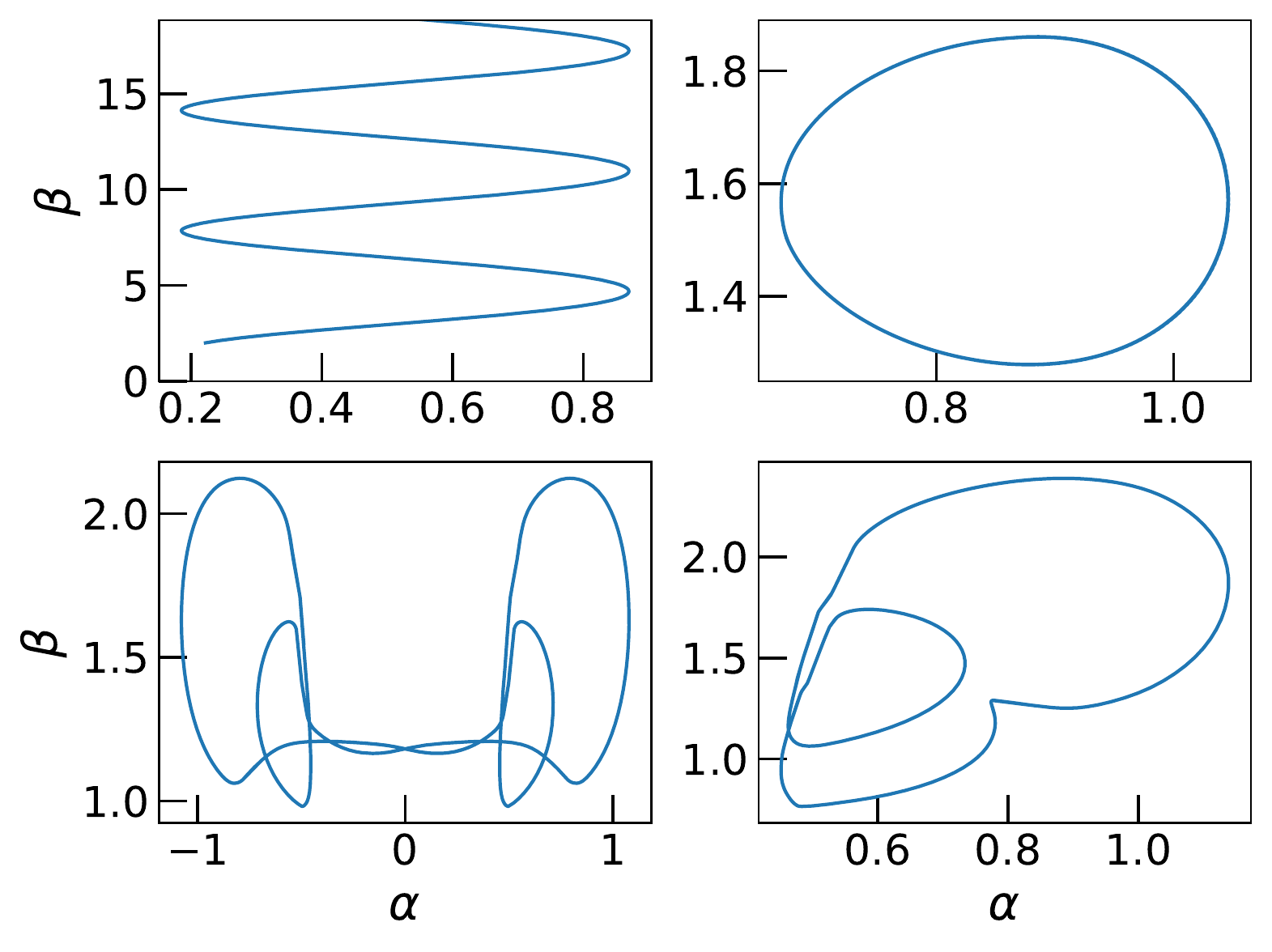}
\caption{Four different long time periodic solutions of Eq.~\ref{eq:quat1} for $\tilde \omega>\tilde \omega_c$, $k=0.614$, $s=0.94$, $G_m=0.01$ of the asynchronous motion projected in  the $\alpha$-$\beta$ plane. The top row correspond to back-and-forth motion and precession obtained by solving Eq.~\ref{eq:quat} with $G_m=0.01$ and $\tilde \omega=1.5$ with the same initial conditions as in Fig.~\ref{fig:PhasePlot2}. In the bottom row the long-time solution's  of Eq.~\ref{eq:quat} with $\tilde \omega=0.9$, $G_m=0.86$ and $G_m=0.94$ projection to $\alpha$-$\beta$ plane are shown. Note as we present projections, trajectories can cross themselves. The motion of the cube for these modes can be seen in Video3 \cite{Video0}.}
\label{fig:gs3}
\end{figure}

In the case when $\tilde \omega>\tilde \omega_c$ there are no new stationary points except $P_3$ and $P_4$, which as follows form Fig.~\ref{fig:PhasePlot2} still remain neutrally stable. This is a bit unexpected as a small parasitic static magnetic field which is perpendicular to the rotating magnetic field leads to a situation where neutrally stationary points become stable \cite{Palkar2019}. As one can see in Fig.~\ref{fig:gs3}, back-and-forth motion and precession are still observable  (top row) as in the case without gravity. However, the oscillation amplitude for angle $\alpha$ is now larger as figures on the top row are obtained from the same initial conditions as the red and green trajectory in Fig.~\ref{fig:PhasePlot2}. There only values of gravity parameter $G_m$ from $G_m=0$ to $G_m=0.01$ were changed. Increasing the parameter $G_m$ even more leads to a larger amplitude for $\alpha$. At some point precession becomes unstable and we observe more complicated 3D motion instead (bottom row in Fig.~\ref{fig:gs3}). Initially a cube rotates on the face. The lag increases, but instead of back motion to catch up with the magnetic field, the cube rolls and trough rotation in third dimension catches up with the  magnetic field.  In fact this motion is a blend of back-and-forth motion and precession (for better understanding different modes reader is advised to examine Video3 \cite{Video0}).
There are two different modes of this complicated dynamics where the magnetic moment rotates around both fixed points or only around one (left and right figure in the bottom row of Fig,~\ref{fig:gs3} respectively). The left mode is observed for smaller values of $G_m$.

\subsection{Two particles}

For the two-particle case the full system of equations has to be solved Eq.~\ref{eq:quat1} and the solution depends on control parameters $s$, $k$, and $G_m$. A 14 dimensional quantity has to be evolved in time. There are four different long-time-limit scenarios possible. If the hematite particles are sufficiently far they rotate independently as described in previous subsection. The other possibilities are that cubes form a stable chain,  or an asymmetric-chain where particles have different types of motion, e.g. one particle rotates on an edge, but other on a face (see Video4 \cite{Video0}). The fourth possibility is that particles undergo motion, where at times a chain is formed which then breaks. The long-time trajectories, in the last two cases, strongly depend on initial conditions.

In a static external magnetic field two hematite cubes form a straight chain \cite{Brics2022}. If the external magnetic field is larger than $\approx0.1$ mT (magnetic field is perpendicular to the gravity) then the magnetic moments of cubes are parallel to the external magnetic field. In this case cubes lie on an edge and the tilt angle (the angle between the base and cube's face) is close to 45$^\circ$ \cite{Philipse2018,Brics2022}. In a very slowly rotating magnetic field such  a chain should rotate with almost no lag. By increasing the rotational frequency the configuration changes and the lag increases. Unlike a single cube, a two-cube chain can not rotate on a corner, due to geometric restrictions.
Note that for a two-cube chain, if it rotates on an edge, the magnetic moment does not have to be in the plane of the magnetic field, as the reaction torque between particles can balance a component of the magnetic torque. However, for values of gravity parameter $G_m$ used in our experiment the magnetic moment is in the plane of the rotating magnetic field. For a two-cube chain also  asynchronous  motion is observed. But there may exist a frequency interval where the chain undergoes disassembly and reassembly motion. This is the case when chain breaks with increase of rotation frequency before back-and-forth motion can happen.

\begin{figure}[ht]
\includegraphics[width=0.5\columnwidth]{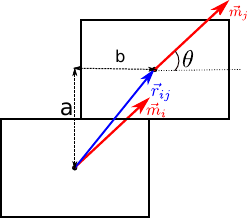}
\caption{Schematic view of two-particle chain from above.}
\label{fig:2cubes}
\end{figure}

 For two-cube chains also  in asynchronous regime not all modes of a single cube motion are present due to geometric restrictions.
 We do not observe precession of the magnetic moment. Due to the same reason for back-and-forth motion the oscillations of the angle $\alpha$ are suppressed whenever magnetic moments of individual cubes are synchronised. Surprisingly, we still observe complicated 3D dynamics where initially the chain rotates and lag increases, but instead of back motion to catch up with the magnetic field, the chain rolls and catches up with the magnetic field through rotation in the third dimension. But unlike for a single cube this mode is only stable  for small values of $G_m<0.15$. For this mode not only the magnetic moment goes out of plane, but cubes are periodically on top of each other (see Video5 \cite{Video0}). For referring later we call it asynchronous out of plane rotation. This mode also should be experimentally easily distinguishable  as the chain's length visible in a microscope changes significantly (ratio of maximal to minimal length is almost two). For $G_m>0.15$ this mode is not anymore observable. Then one observes back-and-forth motion of asymmetric-chain where one cube rotates more on an edge, but other more on a face (see Video4 \cite{Video0}).  
 
If we consider only experimental conditions ($B_{exp}\in [0.3; 3]\,\mathrm{ mT}$ and $a_0\approx1.5\,\mathrm{\mu m}$) then for a two-cube chain there are four regimes of motion. Cube motion for those modes is shown in Video6 \cite{Video0}. Three of them are planar motion regimes: solid-body motion, back-and-forth motion, and periodic chain disassembly and reassembly. For the fourth regime the cube-chain goes out of the plane of magnetic field. During motion the chain rotates slower than the rotating magnetic field and,  in order to catch up with the magnetic field, it rolls on an edge and through rotation in the third dimension  catches up with the magnetic field. This mode is a blend of back-and-forth motion and precession. To give insight into how this motion should look from an experimental perspective the reader can examine Video7 \cite{Video0}, where cubes have been lifted to make shadows better visible. 

\subsubsection{Planar motion with $\alpha=0$}

For planar motion, as long as two cubes form a chain, moments of cubes are synchronized $\theta_1=\theta_2=\theta$. Therefore, the chain  can  be effectively described with two parameters: angle $\theta$ and { horizontal displacement of cube's centers --- shift} $b$ which are defined in Fig.~\ref{fig:2cubes}. 
 
 The EOM in this case is more complicated than for a single particle without gravity, but can easily be obtained using Lagrange mechanics with Rayleigh dissipation function $G(\vect{v},\vect{\Omega})=\frac{1}{2}\sum_i(\xi \vect{v}_i\cdot\vect{v}_i+\zeta \vect{\Omega}_i\cdot\vect{\Omega}_i)$ approach. However, the actual expressions in the general case are quite lengthy, thus they are not provided explicitly. For the case when there exists a stable stationary solution for $b$, one obtains that there is a stationary solution for the lag $\beta= \tilde \omega \tilde t - \theta$
\begin{equation}
 \dot \beta=\tilde \omega\left(1+\frac{1+\tilde b^2}{4k} \right)-\sin(\beta), \label{eq:lag_2p}
\end{equation}
where $\tilde b=b/a_0$.

The critical frequency $\tilde \omega_{c}$ above which no stationary solution for the lag angle may exist is
\begin{equation}
 \tilde \omega_{c}=\frac{4k}{4k +1+\tilde b^2}<1.
 \label{eq:omega_crit_1}
\end{equation}
Note that below critical frequency there still may exist a region where a chain disassembly and reassembly motion is observed. Thus, two additional frequencies $\tilde \omega_s$ and $\tilde \omega_a$ are introduced. The $\tilde \omega_s$ is the maximal frequency until which the chain rotates synchronously with the external magnetic field and $\tilde \omega_a$ is the minimal frequency when asynchronous back-and-forth motion is observed. When $\tilde \omega_s\neq \tilde \omega_a$ this means that there is a region where a chain disassembly and reassembly motion is observed. If $\tilde \omega_s= \tilde \omega_a$ $\Rightarrow$ $\tilde \omega_s=\tilde \omega_{c}= \tilde \omega_a$.

\begin{figure}[ht]
\includegraphics[width=0.9\columnwidth]{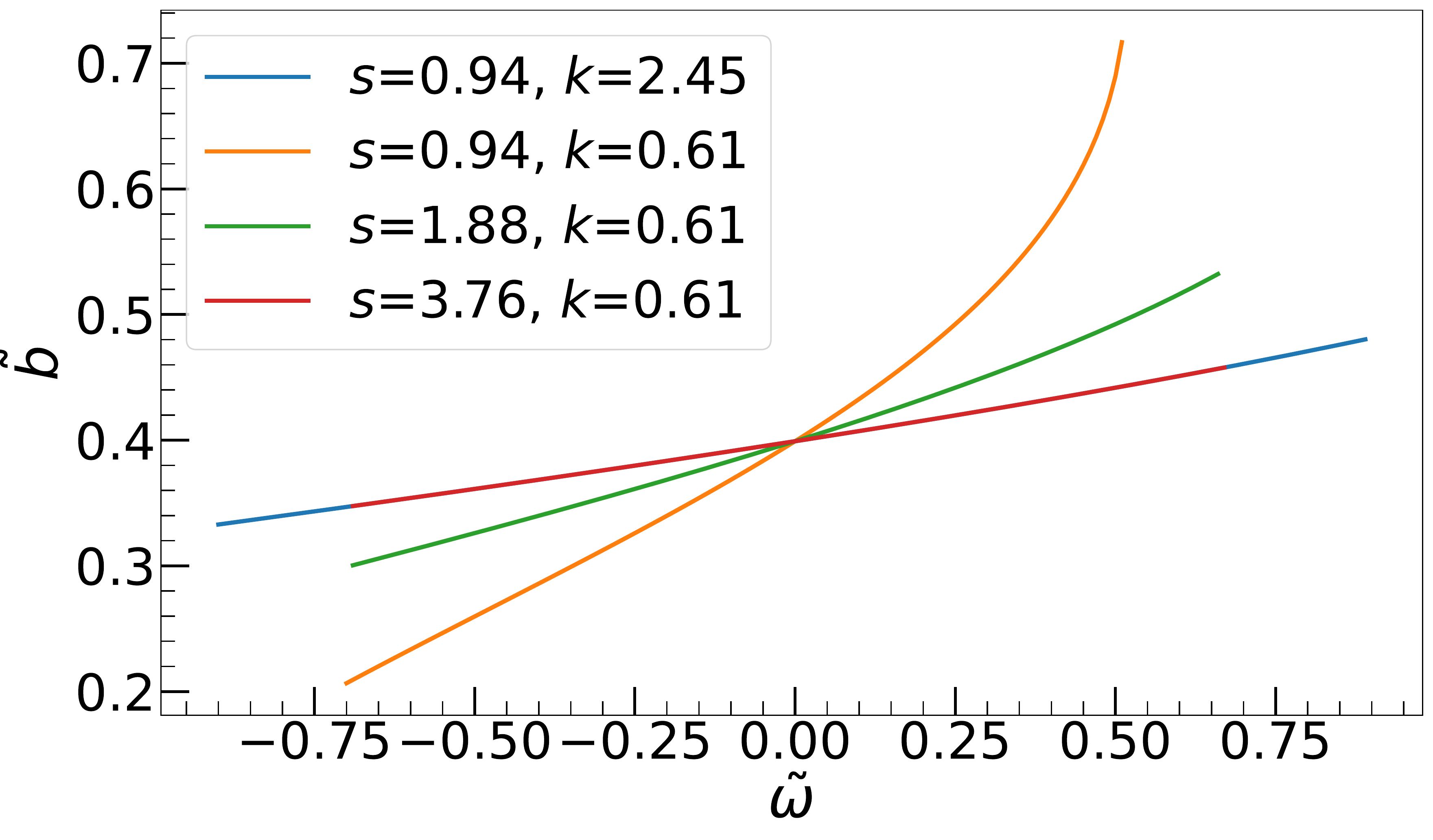}
\caption{The horizontal cube center displacement $\tilde b$ vs. rotation frequency $\tilde \omega$ for a two-cube chain, calculated using Eq.~\ref{eq:quat}.  The shape of the curve is determined by the product of $s$ and $k$, however, the $\tilde \omega_{c}$ according to Eq.~\ref{eq:omega_crit_1} depends on both parameters $s$ and $k$. Thus, red and blue curves overlap, but the blue curve extends further as $\abs{\tilde\omega_c}$ is larger for the blue curve.  The positive value of $\tilde \omega$ means counterclockwise rotation while negative corresponds to clockwise rotation.}
\label{fig:2bvsomega}
\end{figure}

\begin{figure}[ht]
\includegraphics[width=0.5\columnwidth]{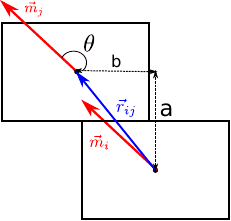}
\caption{Schematic view from above of the second two-particle chain chain configuration. There is a 50\% chance that cubes arrange in this configuration.}
\label{fig:2cubesa}
\end{figure}

For a particular cube chain the critical frequency differs for a clockwise and counterclockwise rotational direction of the magnetic field (depends on the sign of $\tilde{\omega}$). If, for configuration visible in Fig.~\ref{fig:2cubes}, the  rotation speed is increased when  the two-particle chain rotates clockwise, then the shift $\tilde b$ reduces. The opposite, however,  happens when the two-particle chain rotates anti-clockwise as one can see Fig.~\ref{fig:2bvsomega}. This means that for clockwise rotation shift $\tilde b$ is smaller than for anti-clockwise rotation. Therefore, as follows from Eq.~\ref{eq:omega_crit_1},  the critical frequency is larger for the clockwise case. This breaks the symmetry. However, there exists the other two-particle configuration (see fig.~\ref{fig:2cubesa}), which is observable with the same probability \cite{Brics2022}. For this configuration the critical frequency is larger for the counterclockwise case.  This guarantees that on average the observations are the same for both directions of the rotations. Therefore, without loss of of generality, only the configuration shown in fig.~\ref{fig:2cubes} is analyzed. For the second configuration, which is shown in fig.~\ref{fig:2cubesa}, one obtains the same dynamics as for the first  if rotation direction (the sign of $\tilde \omega$) is reversed.

\begin{figure}[ht]
\includegraphics[width=0.48\columnwidth]{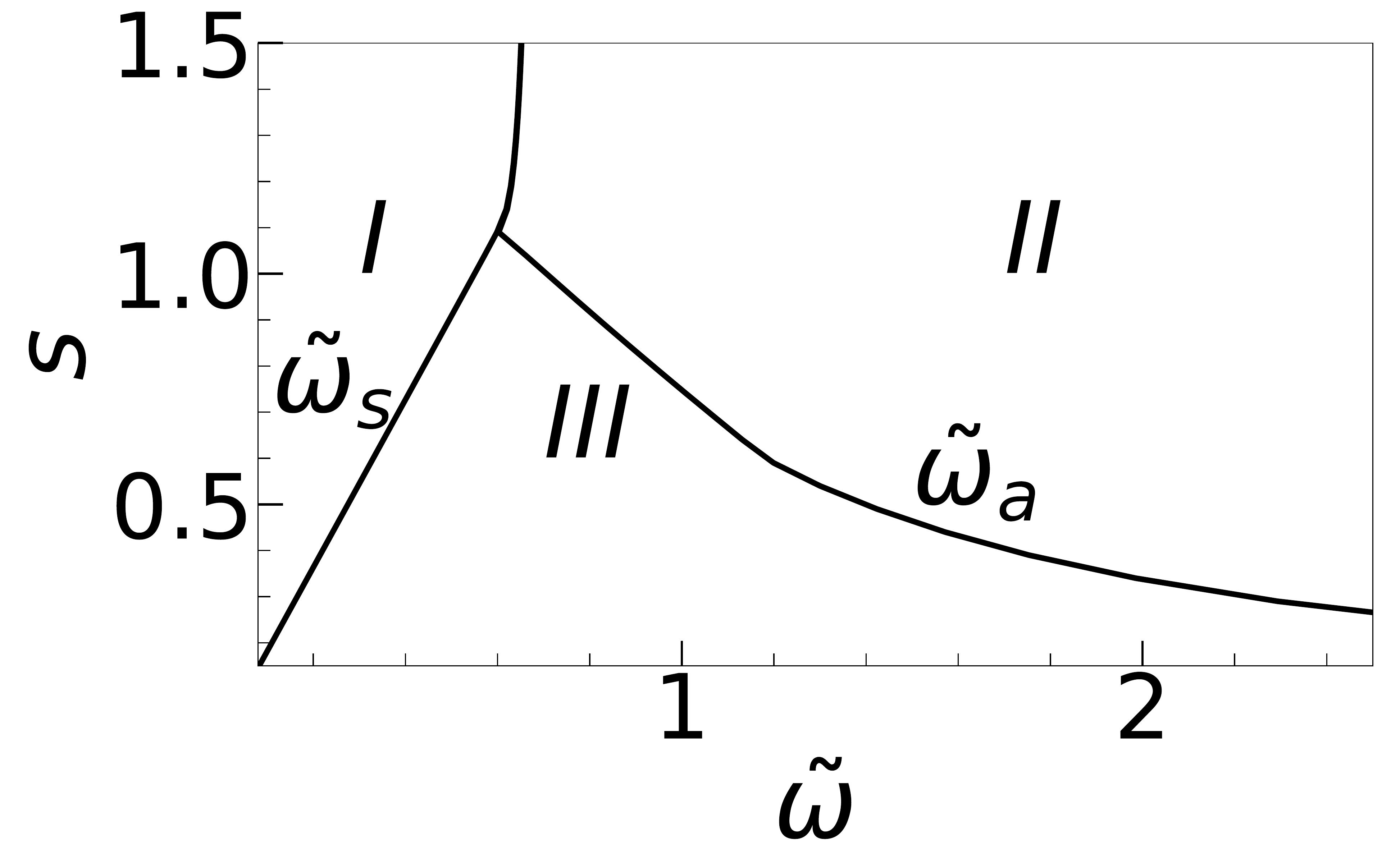}\hfill
\includegraphics[width=0.48\columnwidth]{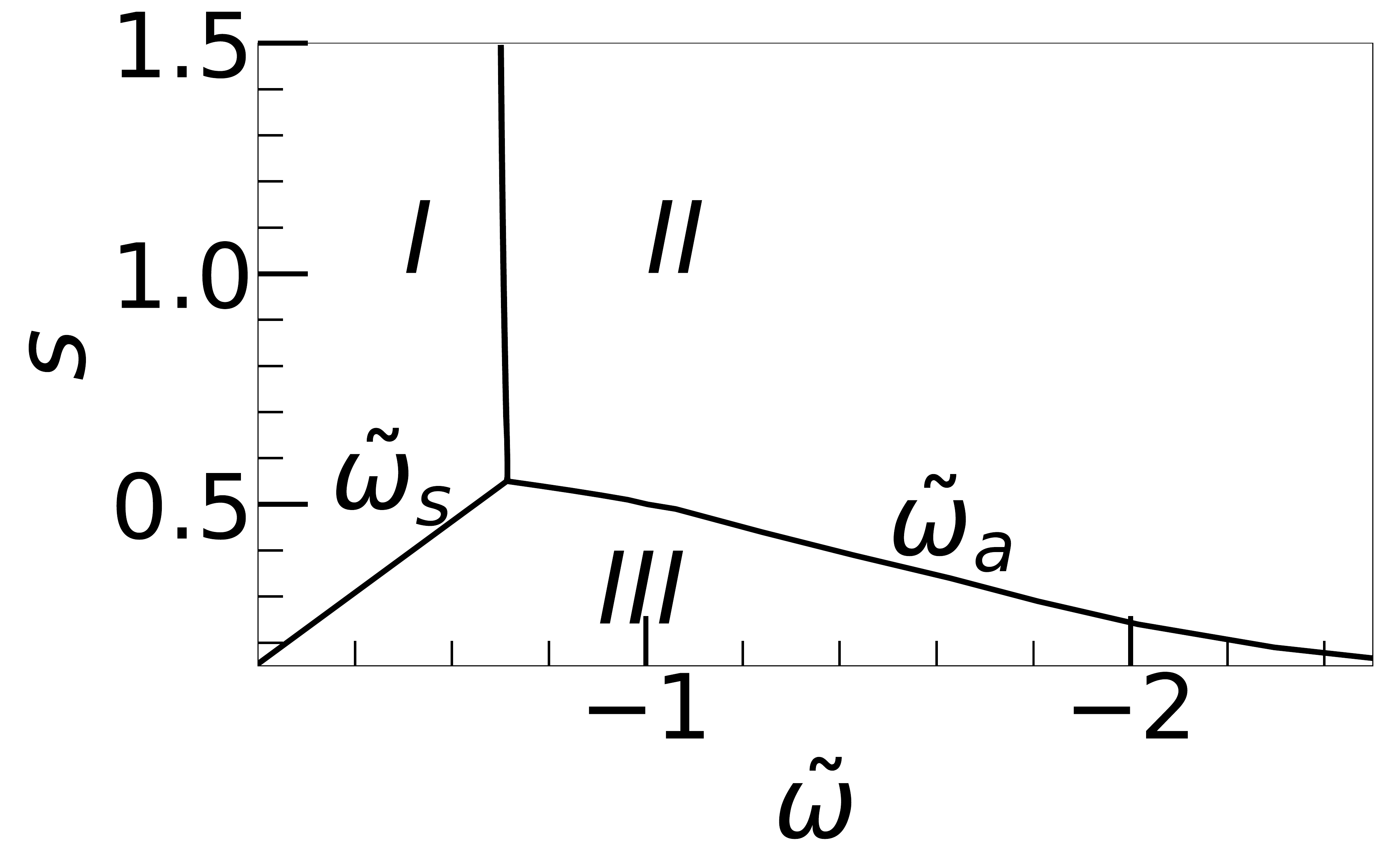}\\
\includegraphics[width=0.48\columnwidth]{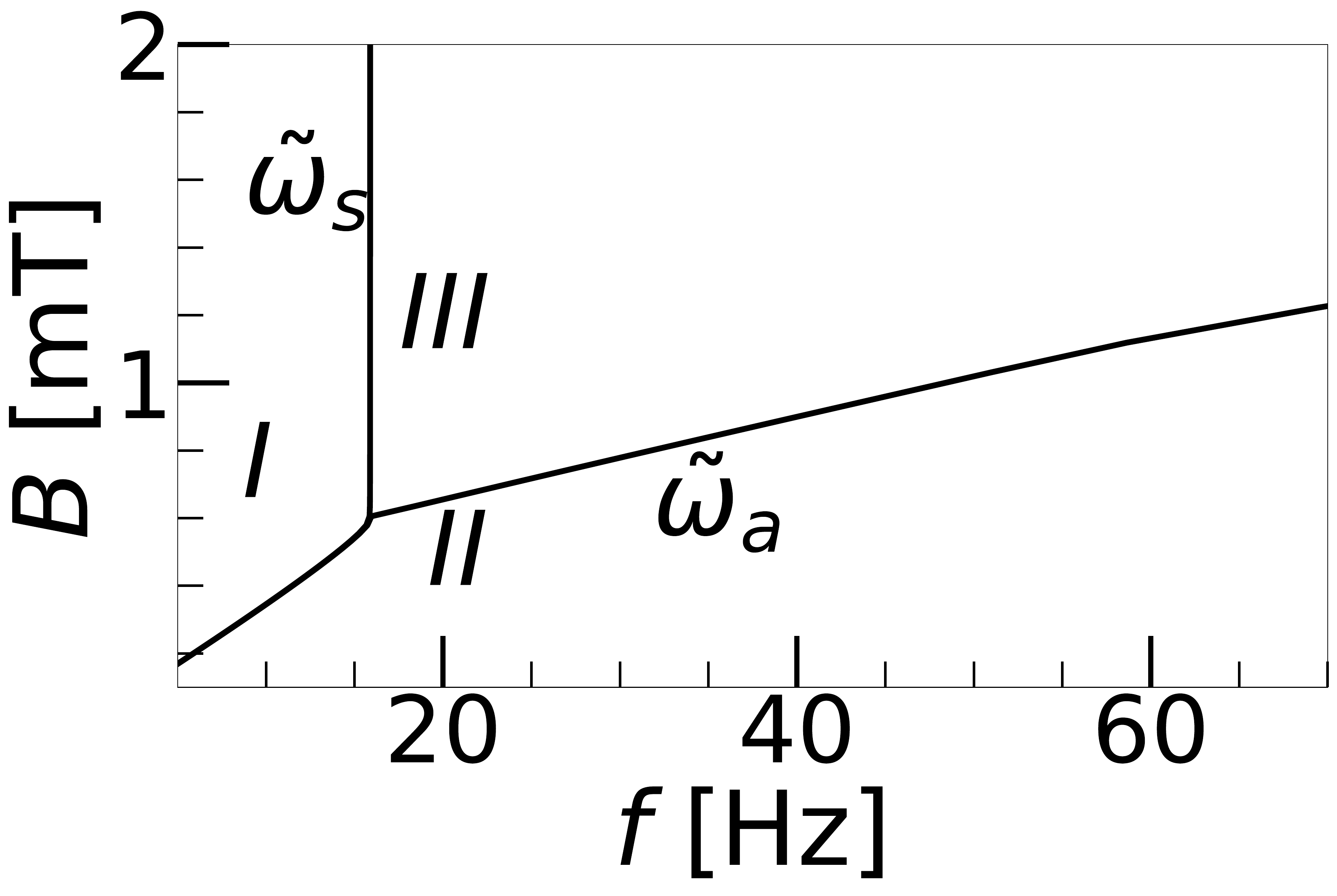}\hfill
\includegraphics[width=0.48\columnwidth]{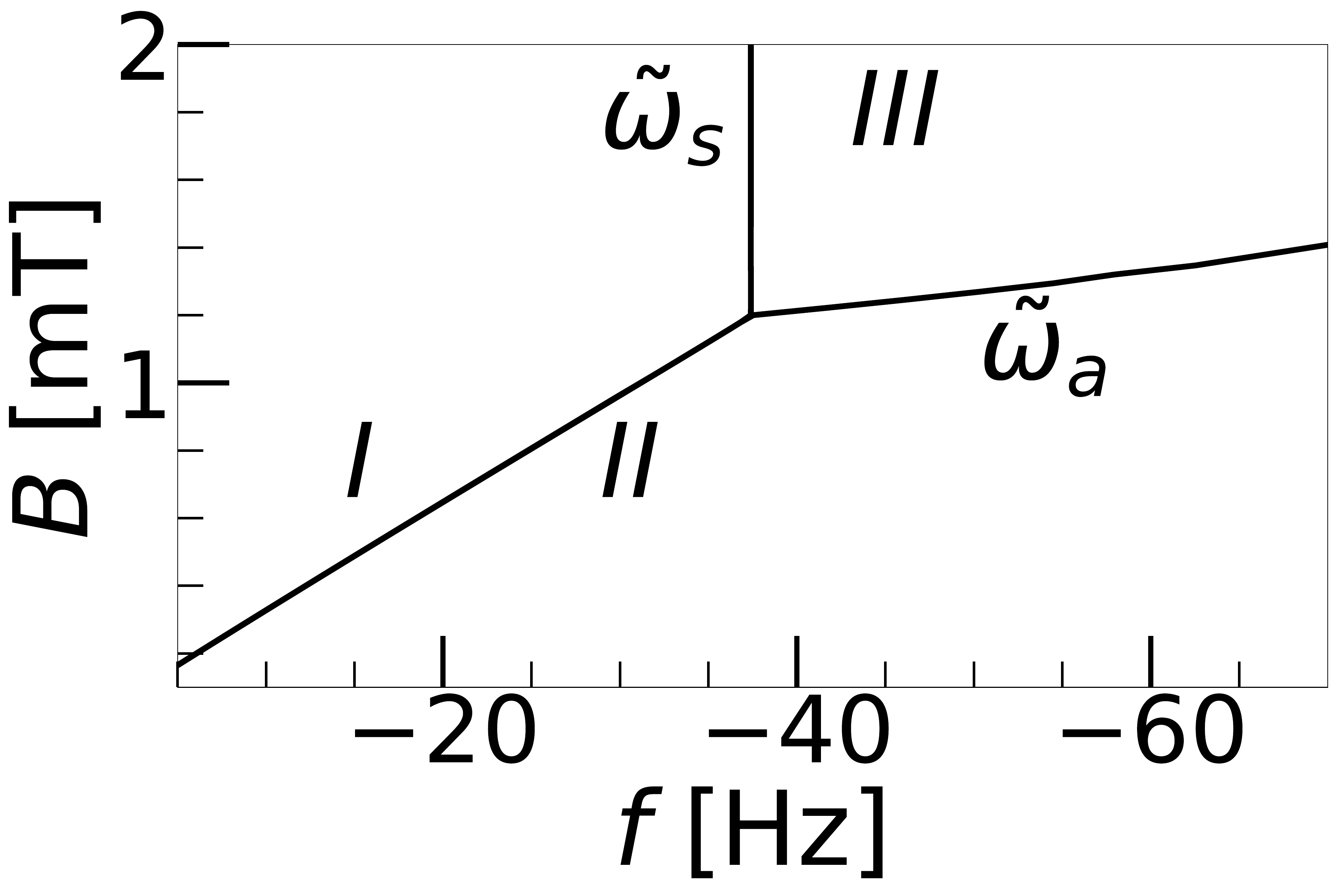}
\caption{Borders of long-time dynamics regimes of Eq.~\ref{eq:quat} for $k=1.84$ using dimensionless variables (top row) and dimensional variables for comparison with experiment (bottom row). The regions I,II, III correspond to solid body rotation, back-and-forth motion, periodic chain disassembly and reassembly motion regimes respectively.  The positive value of $\tilde \omega$ means counterclockwise rotation while negative corresponds to clockwise rotation.}
\label{fig:stability}
\end{figure}

\begin{figure}[ht]
\includegraphics[width=0.9\columnwidth]{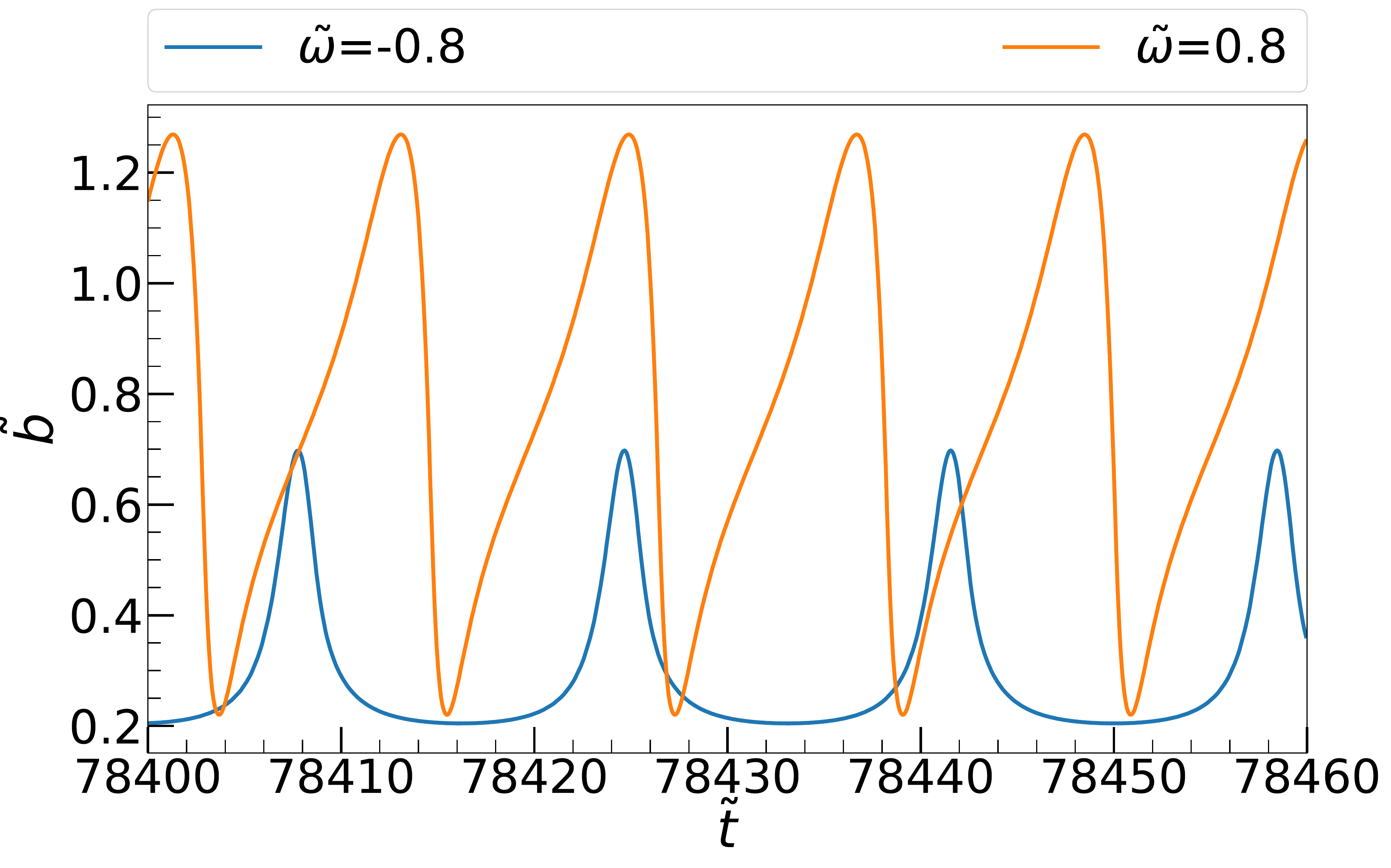}
\caption{The time evolution after transition time of horizontal cube center displacement $\tilde b$ calculated using Eq.~\ref{eq:quat} with $s=0.94$, $k=1.84$, and $\abs{\tilde \omega}=0.8$.  The positive value of $\tilde \omega$ means counterclockwise rotation while negative corresponds to clockwise rotation.}
\label{fig:2bvst}
\end{figure}

Overall, if two particles are not too far apart, there are three regimes for long-time dynamics. For value of $k=1.84$ they are shown in Fig.~\ref{fig:stability}. There in the region I ($|\tilde \omega|<|\tilde \omega_s|$), chain rotates as a solid object with the frequency of rotating field. In the region II ($|\tilde \omega|>|\tilde \omega_a|$) the back-and-forth motion is observed. In this case the shift $\tilde b$ is not anymore constant, but periodically oscillates. The  trajectories in the case of $s=0.94$, $k=1.84$, and $\abs{\tilde \omega}=0.8$ are shown if Fig.~\ref{fig:2bvst}. The oscillation amplitude for $\tilde b$ is always larger for counterclockwise rotation of magnetic field and, in general, reduces by increasing rotational frequency. In the region III ($|\tilde \omega_s|<|\tilde \omega|<|\tilde \omega_a|$   of parameters the chain break. After some time chain can reassemble, but it will break again. We observe a periodic disassembly and reassembly of a chain, thus the long time trajectory is periodic, however, it strongly depends on initial conditions. In the case of positive $\tilde \omega$ as thermal fluctuations are present in the experiment, it can happen that during this motion cubes rearrange in the second configuration (Fig.~\ref{fig:2cubesa})  as in this case the second configuration is energetically favorable.     

\subsection{More than two particles}

\begin{figure}[ht]
\includegraphics[width=0.9\columnwidth]{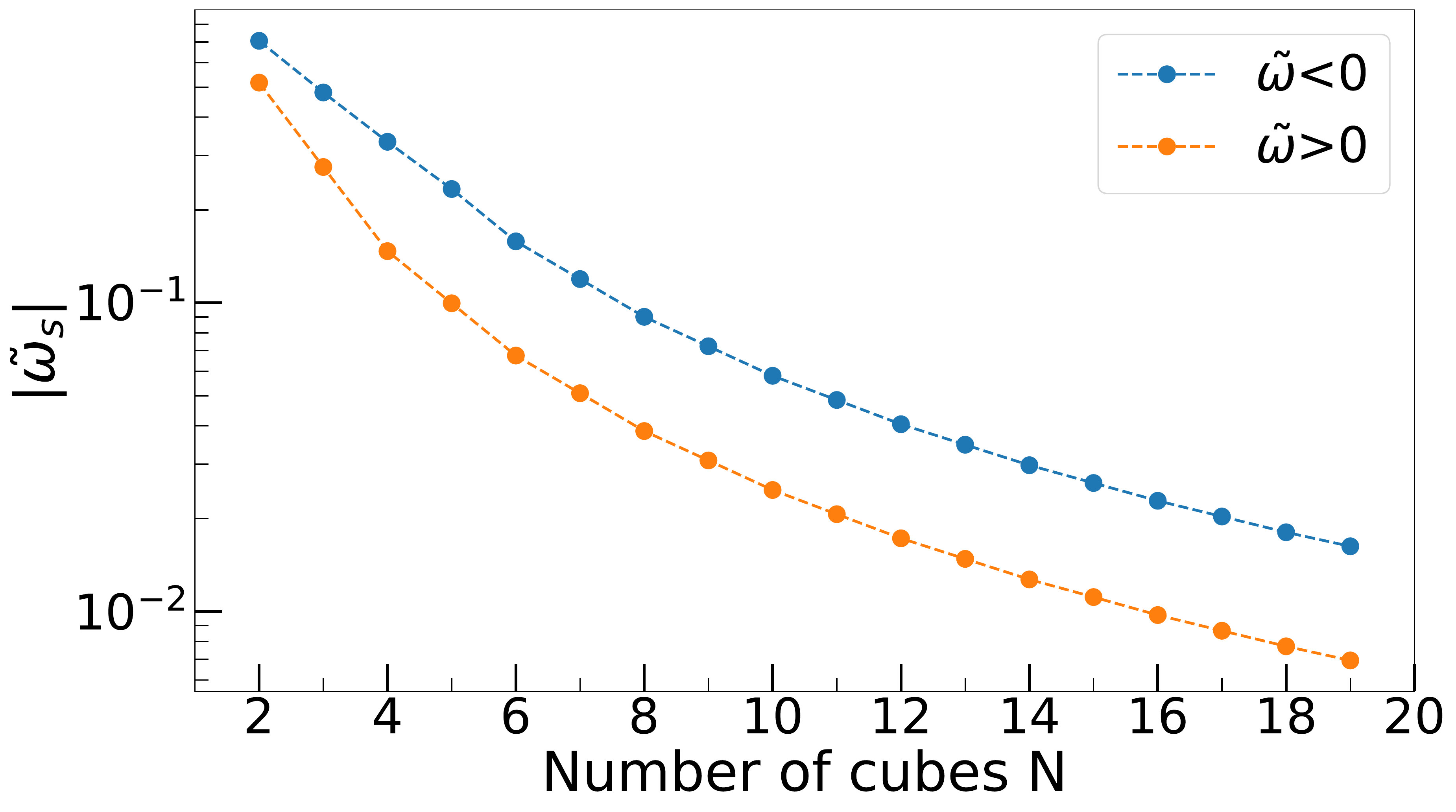}
\includegraphics[width=0.83\columnwidth]{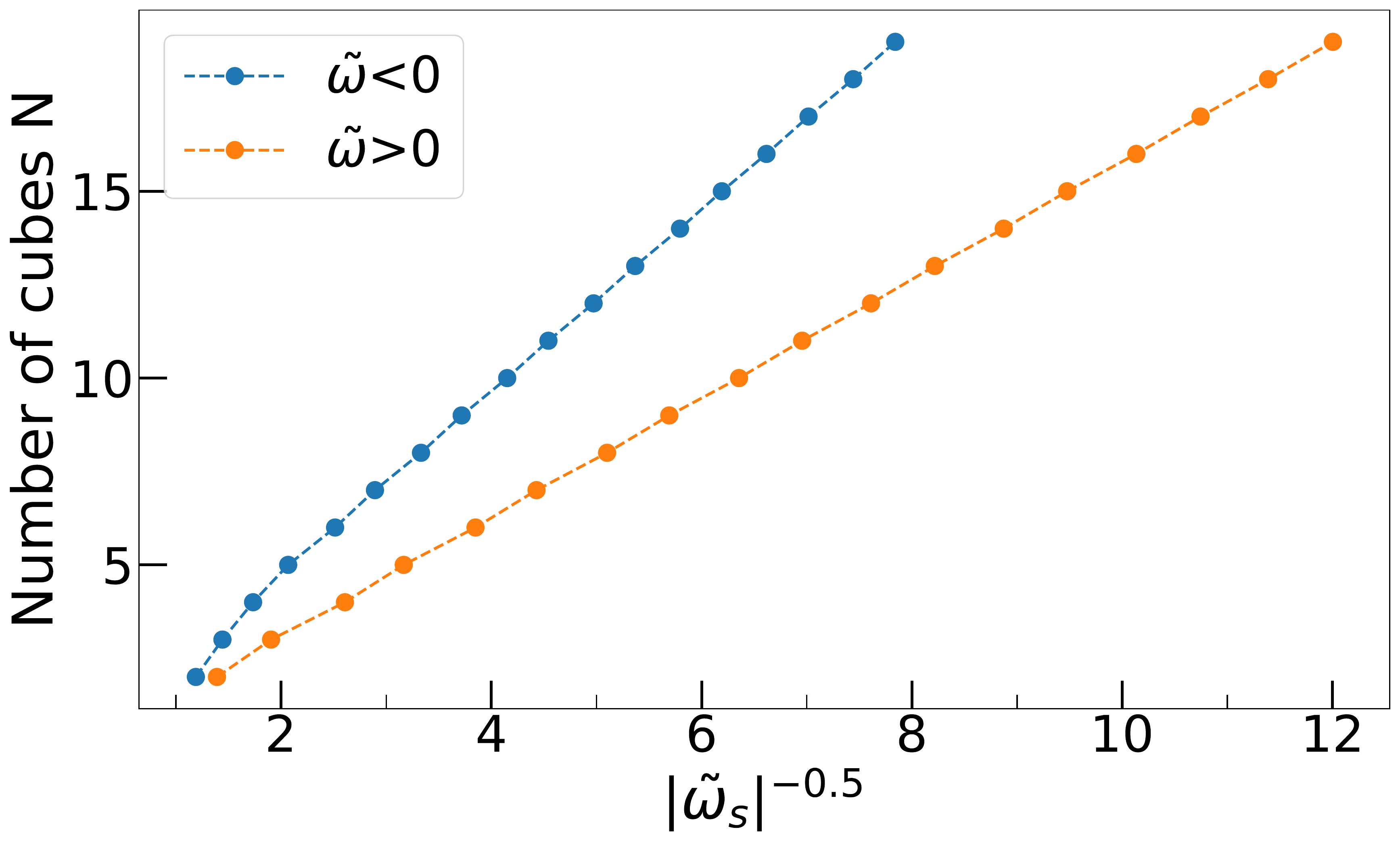}
\caption{Dependence of maximal rotation frequency $\abs{\tilde\omega_s}$ when chain rotates as solid object vs number of hematite particles in a chain. The points are calculated from Eq.\ref{eq:quat} with $s=0.94$ and $k=1.84$. The positive value of $\tilde \omega$ means counterclockwise rotation while negative corresponds to clockwise rotation.}
\label{fig:omega_crit}
\end{figure}

\begin{figure}[ht]
\includegraphics[width=0.32\columnwidth]{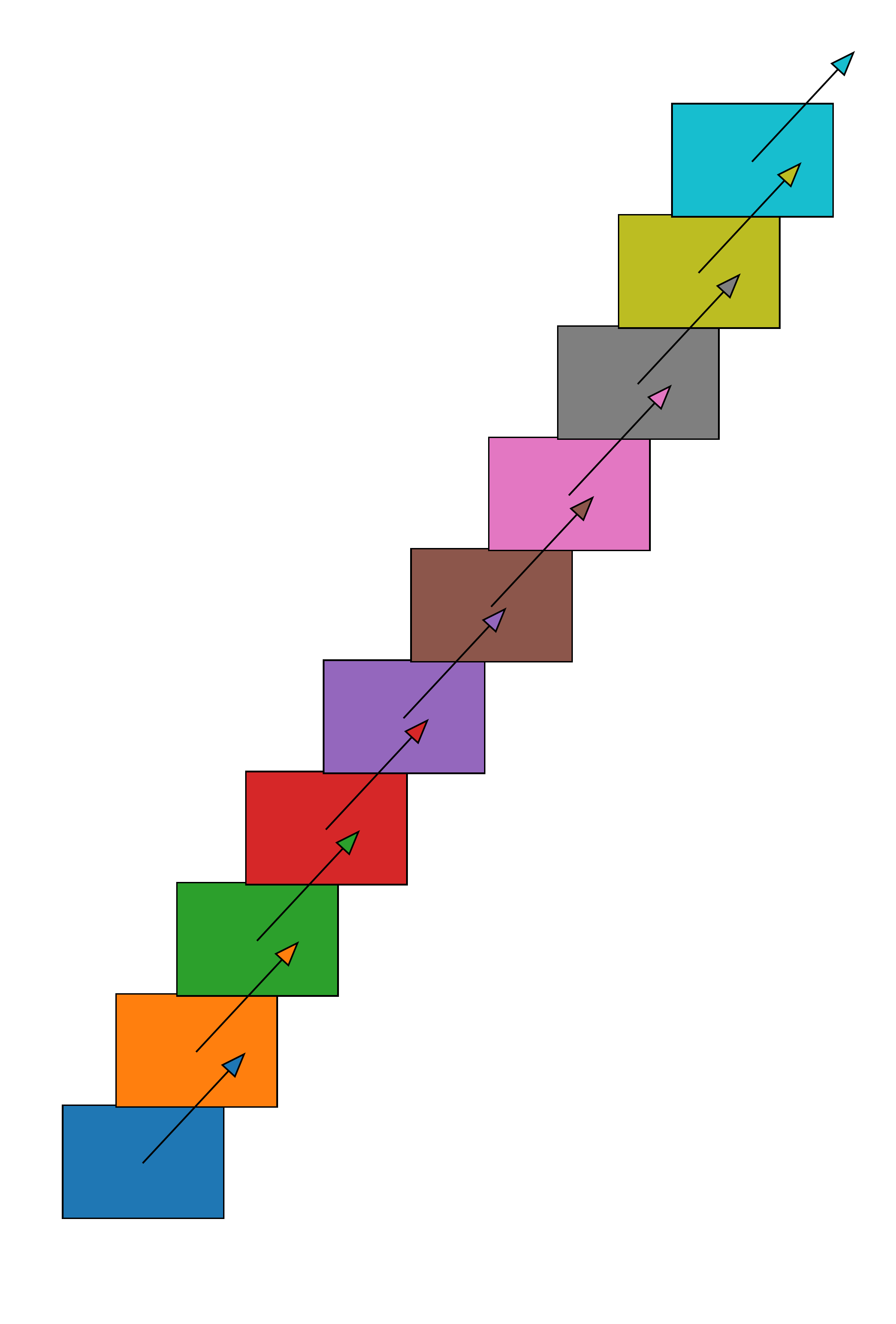}\hfill
\includegraphics[width=0.32\columnwidth]{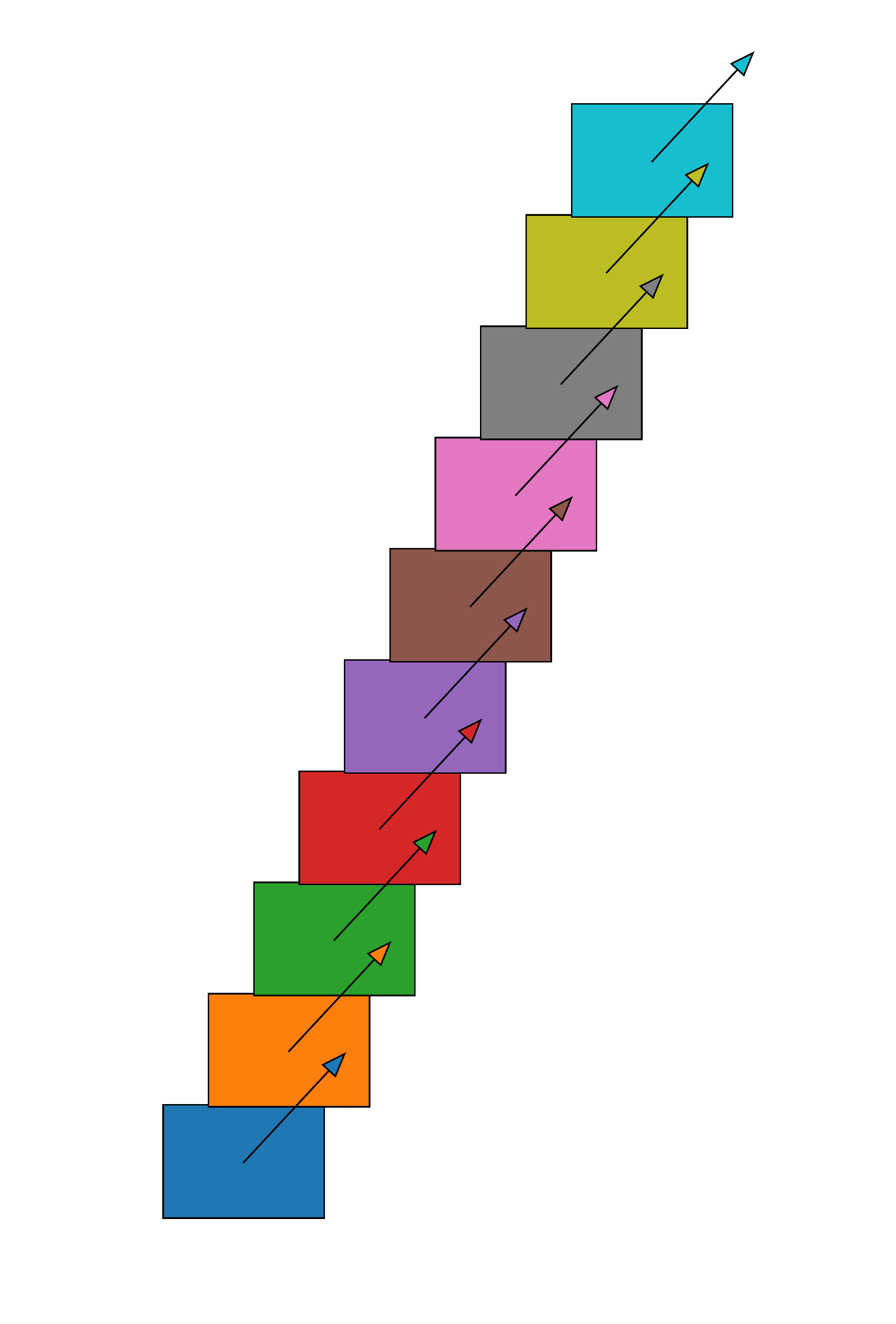}\hfill
\includegraphics[width=0.32\columnwidth]{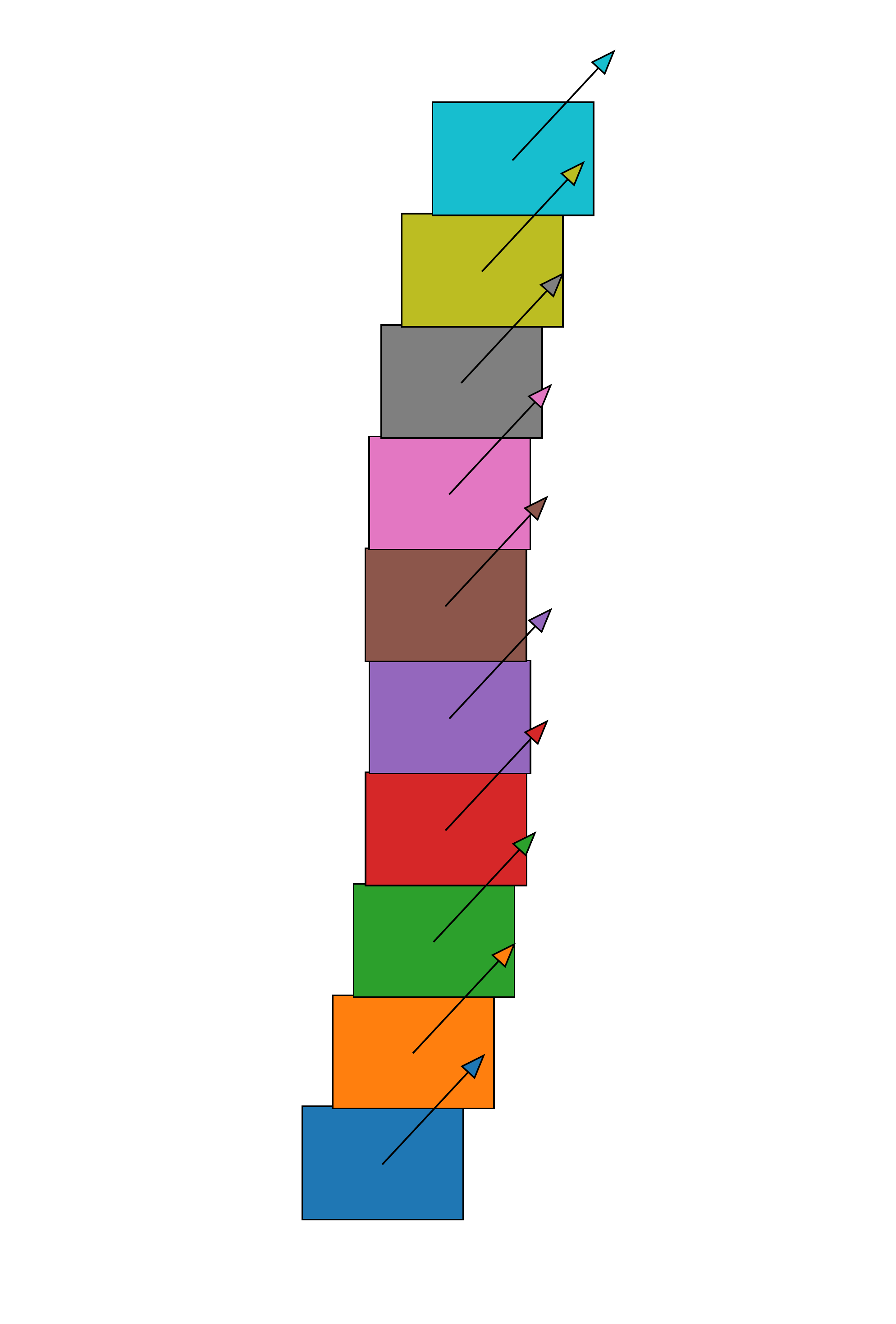}
\caption{Configuration of ten particle chains (from left to right),  which is in an anti-clockwise rotating magnetic field, a static magnetic field, and a clockwise rotating magnetic field. The configurations are obtained from Eq.\ref{eq:quat} with $s=0.94$ and $k=1.84$. In the case of the rotating magnetic field, the frequency $\abs{\tilde\omega}$ is just below $\abs{\tilde\omega_s}$. If $\abs{\tilde\omega}$ is increased the chains break. }
\label{fig:10_part}
\end{figure}

In this case, depending on how particles are distributed, several chains may be formed which undergo more or less independent motion. When a single chain is formed there are the same three planar motion regimes. The regime where the cube chain goes out of the plane of rotation  magnetic field is observable for chains consisting of up to four particles. For planar motion qualitatively we obtain the same diagram as in Fig.~\ref{fig:stability} only the region III increases and both region I and region II become smaller with increasing number of particles. For the fixed values of $s$ and $k$ the $\abs{\tilde \omega_s}$ decreases with increasing number of particles $N$ in the chain, as shown in Fig.~\ref{fig:omega_crit}. For maximal chain length which is observed we obtain that $N\propto \frac{1}{\sqrt{\abs{\tilde \omega}}}$. This is the same relation as for paramagnetic spherical particles with and without anisotropy \cite{Melle2003,Drikis2004, Uzulis2022}. {Concerning chain shape a similar effect to bending of chains of spherical paramagnetic particles \cite{Melle2003,Drikis2004, Biswal2004, Javaitis2004, Zaben2020, Uzulis2022} is observed, but the shape is quite different and depends on rotation direction (see Fig.~\ref{fig:10_part})}.

The opposite behavior is observed for $\abs{\tilde \omega_a}$, thus also region II becomes smaller with increasing number of particles in a chain $N$. The value of $\abs{\tilde \omega_a}$ for larger $N$ becomes very large. Therefore it is expected that for larger chains the asynchronous regime is not observed experimentally as we are far away from the validity region of our model. 

As for two-particle chains, also here we observe that for a particular chain the behavior depends on rotation direction. For a ten-particle chain shown in Fig.~\ref{fig:10_part} we observe that critical frequency $\abs{\tilde \omega_a}$ is more than two times larger in the case of clockwise rotation. Visually the configuration also looks different. However, when a chain breaks, it always breaks in the middle. If there are an even number of particles in the chain then the chain breaks into two chains with $\frac{N}{2}$ particles in each. If there is an odd number of particles in the chain, it breaks into 3 parts with $\frac{N-1}{2}$, 1, and $\frac{N-1}{2}$ particles in each. {The long-time solution, as for two-particle chains, is a periodic disassembly and reassembly of chain and trajectory is strongly initial condition dependent. However, we do not observe formation of clusters after a chaotic transition period as for paramagnetic particle chains \cite{Abdi2018}.}

\section{Experimental results}
\label{sec:exp}
Behavior of two hematite cube chains in a rotating magnetic field is measured experimentally. For this, hematite cubes are synthesised and characterized following the methods described in \cite{Petrichenko_2020}. The cubes were found to have edge length $a\approx1.5\,\mathrm{\mu m}$ and shape factor $q\approx2.0$. The same sample of hematite cubes was used in all experiments. Mixing of the sample and restoration of chemical composition needed for pH level and suspension stability was done before each experiment.
	
The hematite samples were contained in glass capillaries with  $100\,\mathrm{\mu m}$ thickness filled with liquid.  The observation was done with a microscope (Leica DMI3000B) equipped with a camera (Basler ac1920-155um, up to $250$ frames per second), using an oil immersion objective with $100\times$ magnification. Image acquisition was done with the proprietary camera software while image processing and analysis was performed with MATLAB. In general terms, image analysis relied upon cross correlating an image of a single hematite cube to the experimental image, with the two correlation maximums corresponding to the two cubes of the dimer. After thus identifying the individual cubes, information about their distancing and angle between the axis of the cube chain and the magnetic field could be obtained. These parameters were then used to classify the chain configuration (such as planar motion, periodic chain breakup and reassembly, or out of plane motion) and motion characteristics (correspondence between rotation frequencies of the chain and magnetic field) as belonging to one of the rotation regimes as described in Sec.~\ref{sec:theor_res}.

\begin{figure}[htbp]
\includegraphics[width=0.8\columnwidth]{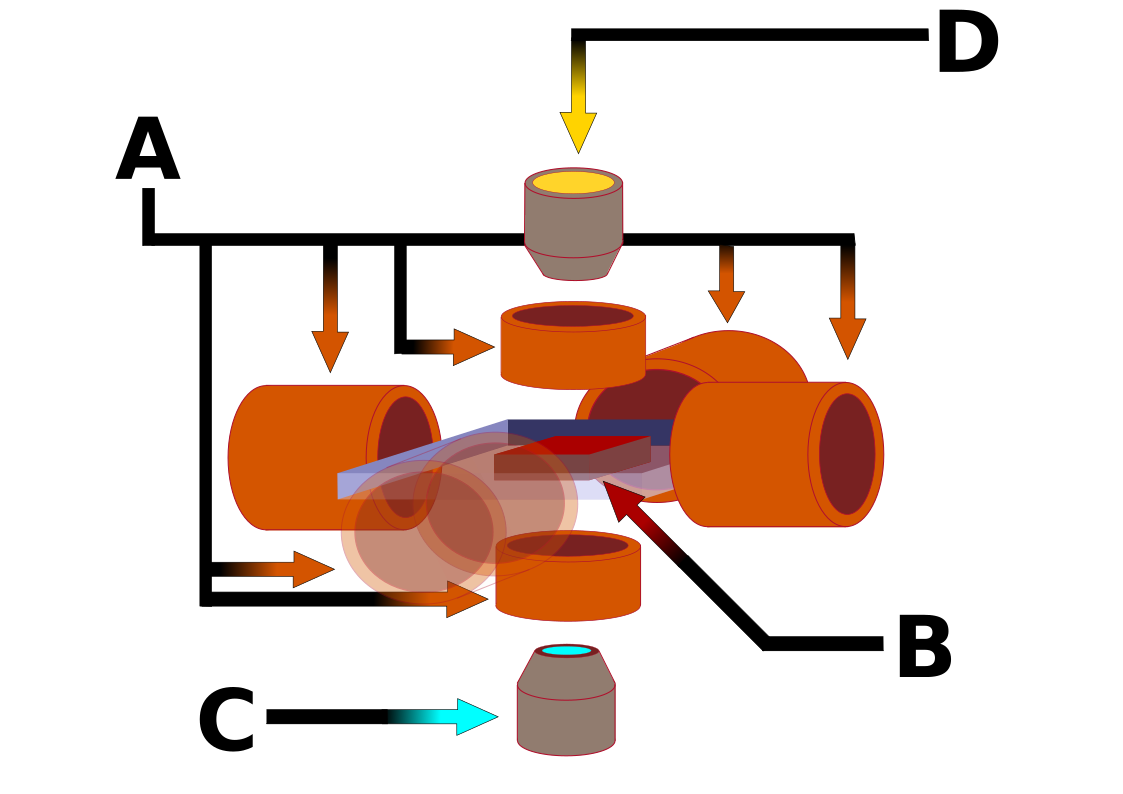}
\caption{Coil system and microscope setup used in experimental work. A – coil system; B – placement of field sensor or sample; C – objective; D – light source}
\label{fig:coil_system}
\end{figure}

The magnetic field was generated by three pairs of coils, powered by DC current sources (KEPCO) that are controlled by a NI DAQ card using LabView code. The glass capillary was placed near the point of crossing of the three coil pair axis (see scheme of layout in Fig.~\ref{fig:coil_system}). To compensate for parasitic magnetic field sources, such as Earth or those associated with lab equipment, the field was measured at the location of the capillary within the microscope. This was done prior to each experiment, using a magnetic sensor (HMC5883 GY-271 3V-5V Triple Axis Compass Magnetometer Sensor Module for Arduino), obtaining background field values via Arduino and then accounting for them in the LabView code. Such a method allows us to define the magnetic field with a precision $\Delta B\in(0.01;0.03)\,\mathrm{mT}$.
	
Hematite cube chain rotation was captured as a sequence of images and accompanying coil current measurements, providing information about the magnetic field. A sequence of measurements for one hematite cube chain would involve either an increase or decrease of field rotation frequency at a constant field magnitude, or a change of magnitude at a constant rotation frequency, or both. Several such sequences are measured for each pair of particles. Magnetic field was increased and decreased continuously, to avoid a stepwise supply of energy to the system. Experimental data presented here was gathered from 49 dimers, and a total of 845 measurements.

	
Several aspects connected to cube chain rotation analysis have to be noted. First, experimentally it is impossible to measure the angle between the external field and magnetic moment. Instead, we use the lag angle between the external field and the axis of the cube chain. Second, initial configuration of the cube chain cannot be set, therefore it has to be determined. We find it by applying a stationary magnetic field and observing the angle between the external field and the axis of the cube chain (see \cite{Brics2022} for more details on the cube chain orientation in stationary magnetic fields). Due to the limited frame rate and optical resolution, this configuration control is repeated not only at the start and end of the measurement series, but also during it. We only use measurements, where the chain configuration has not changed between two control measurements. Third, in experiments we use rotating magnetic fields rotating both in clockwise and counterclockwise directions, regardless of initial configuration of cube chains. This is simpler for the experimental sequence control and allows to check if there are differences in results depending on the initial configuration and rotation direction.

\begin{figure}[ht]
\includegraphics[width=0.85\columnwidth]{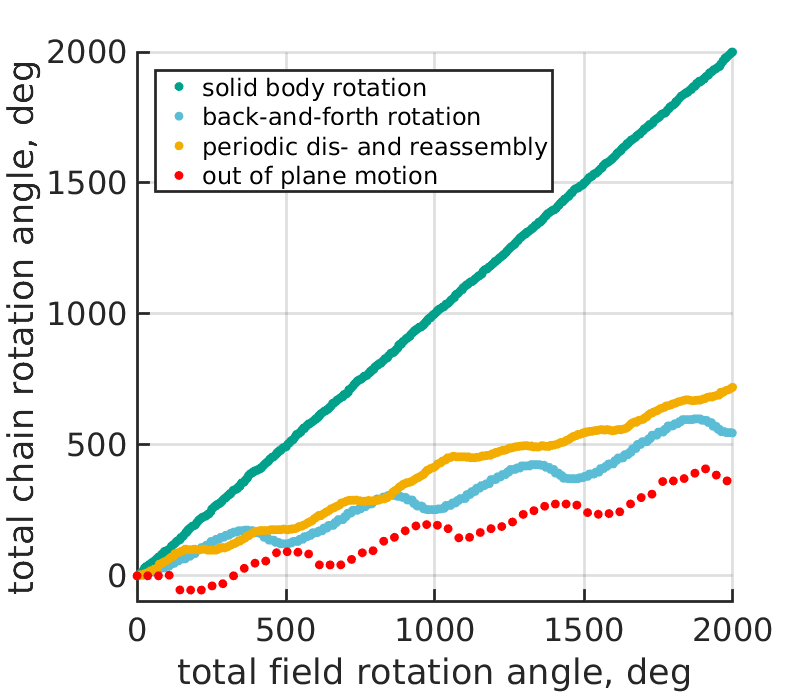}
\caption{Relation between angle of rotation for a hematite cube chain to angle of rotation for magnetic field, in the four regimes of rotation. For a motion out of the field rotation plane, the curve is not continuous as when particles are exactly on top of each other chain rotation angle is not defined. 
} \label{fig:lags}
\end{figure}
\begin{figure}[ht]
\includegraphics[width=0.85\columnwidth]{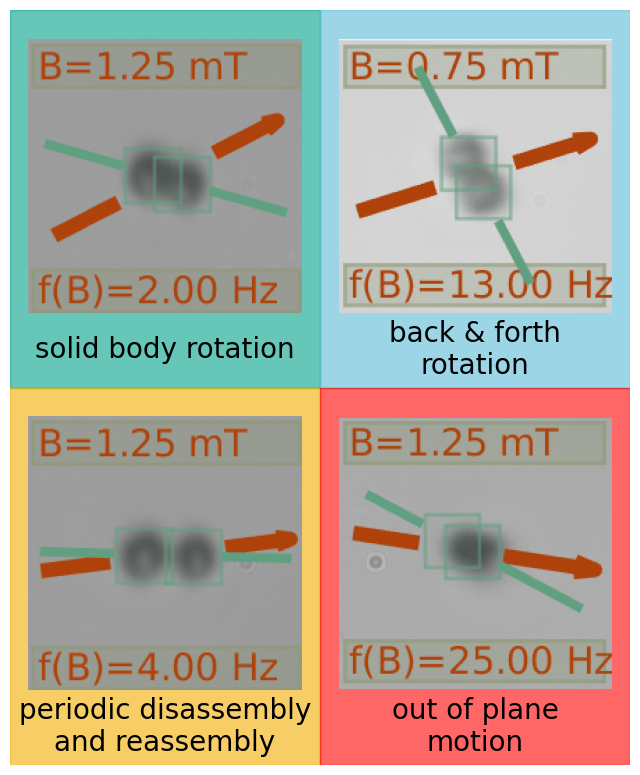}
\caption{Example images from the four rotation regimes. Cube motion measured experimentally can be seen in Video8 \cite{Video0}.} \label{fig:regs}
\end{figure}	
 
An insight into the different regimes of rotation as seen from experimental measurements is provided in Fig.~\ref{fig:lags}. There the relation for angle of rotation for the magnetic field and for the hematite cube chain is provided. 
Based on  experimental videos  and angle measurements we  find the same four different regimes as in theory (see Sec.\ref{sec:theor_res}) of which three are planar rotation and one which is associated with asynchronous rotation where chain goes out of the plane of the rotating magnetic field. From them only the solid-body regime follows the magnetic field, while rotation frequency and therefore the rotation angle is lower for all the other regimes. The solid-body regime is the only one where the chain rotates synchronously with the magnetic field. Example pictures of the chain, along with illustrations of chain axis and field direction, in each of the regimes are given in Fig.~\ref{fig:regs}.

To have a better comparison with theoretical results, the measurements can be split in two groups, corresponding to the two directions of rotation for a cube chain configuration shown in Fig.~\ref{fig:2cubes}, as was done in Sec.~\ref{sec:theor_res}. In practice it means that the measurements of the first initial cube chain configuration with a subsequently applied clockwise field are combined with the measurements of the second initial position and a rotating field rotating counterclockwise (the combined set of data points further referred to as "clockwise equivalent"), and the opposite (the combination called "counterclockwise equivalent").

   	\begin{figure}[ht]
 		\includegraphics{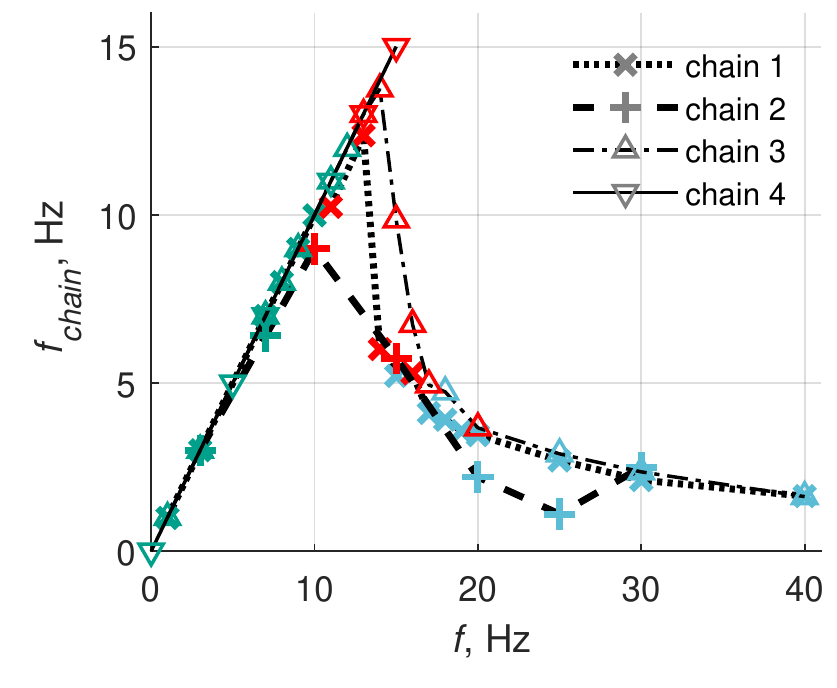}\\
            \includegraphics[width=0.99\columnwidth]{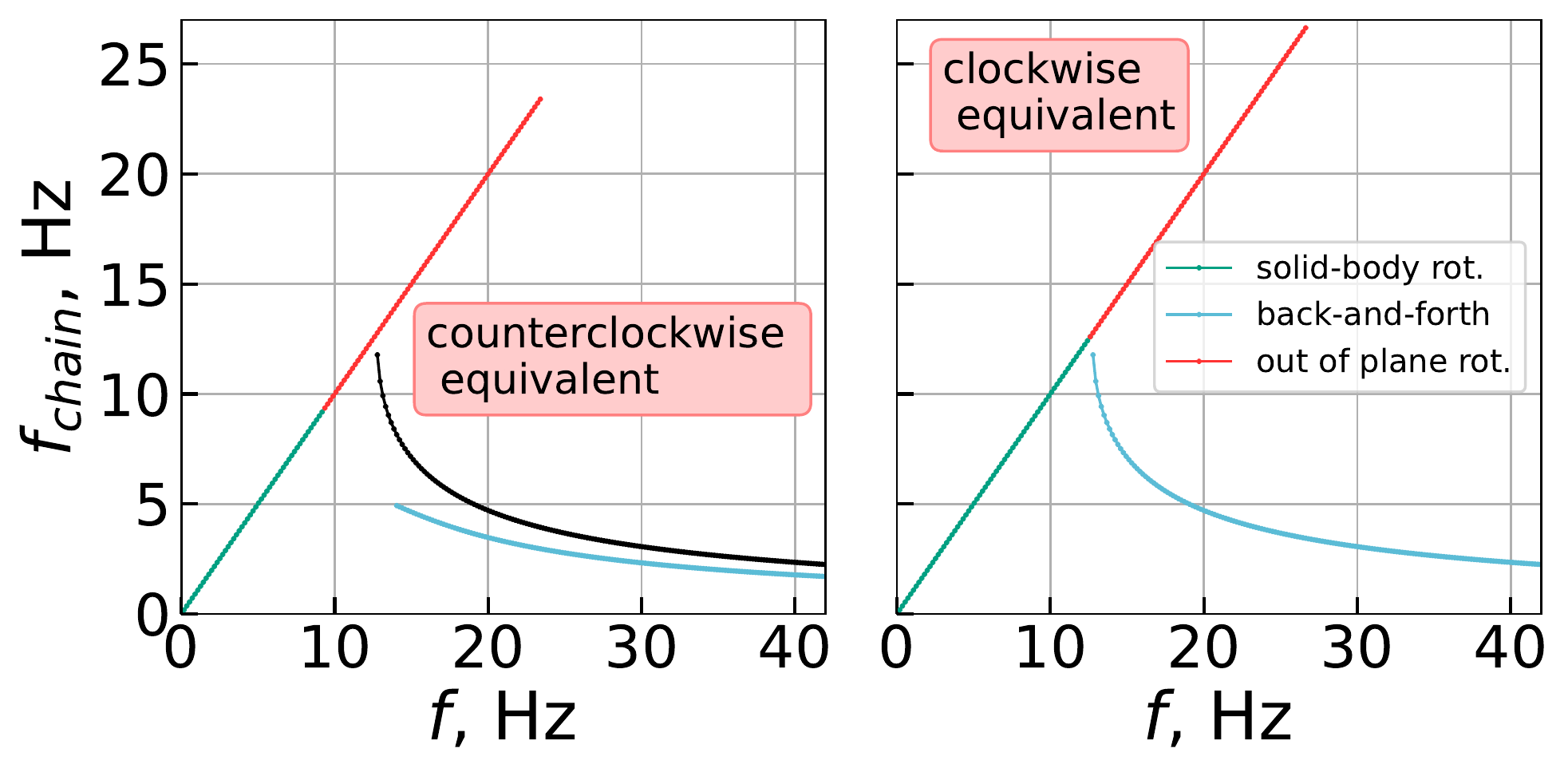}
 		\caption{ Experimentally measured average chain {(N=2)} rotation frequencies for four different two cube chains at various field rotation frequencies, $B=1\,\mathrm{mT}$ (top figure) compared to the simulation results (bottom figures).Experimental data points for each chain are marked with one marker and connected with tracer line. Pluses and crosses are clockwise equivalent; triangles --- counterclockwise equivalent. For theoretical calculations in the counterclokwise equivalent case in the frequency range, where there is no solid body and back-and-forth rotation, periodic dis- and reassembly of chain is observed. The black line is drawn for comparison purposes and shows back-and-forth motion in the clockwise equivalent case.  Colors represent different rotation regimes as in Fig.\ref{fig:regs}.}
 		\label{fig:ffs}
 	\end{figure}

Using experimental data from angle measurements (as shown in Fig.\ref{fig:lags}), we can calculate the average rotation frequency for a chain and show its dependence on magnetic field frequency. In Fig.\ref{fig:ffs} we present experimental results for four different two cube chains at a fixed $B=1\,\mathrm{mT}$, indicated by different markers and tracer lines, along with theoretical predictions. Although the measurement series for each of the chains look similar to classical rotating rod \cite{Erglis2011}, behavior is more diverse. In particular, more rotation regimes are present - out of plane motion (red symbols) and periodic dis- and reassembly (not observed for these four chains) complement solid body (green) and back-and-forth (blue) rotation (same colors as in Fig.\ref{fig:regs}).

\begin{figure}[htbp]
    \includegraphics[width=0.95\columnwidth]{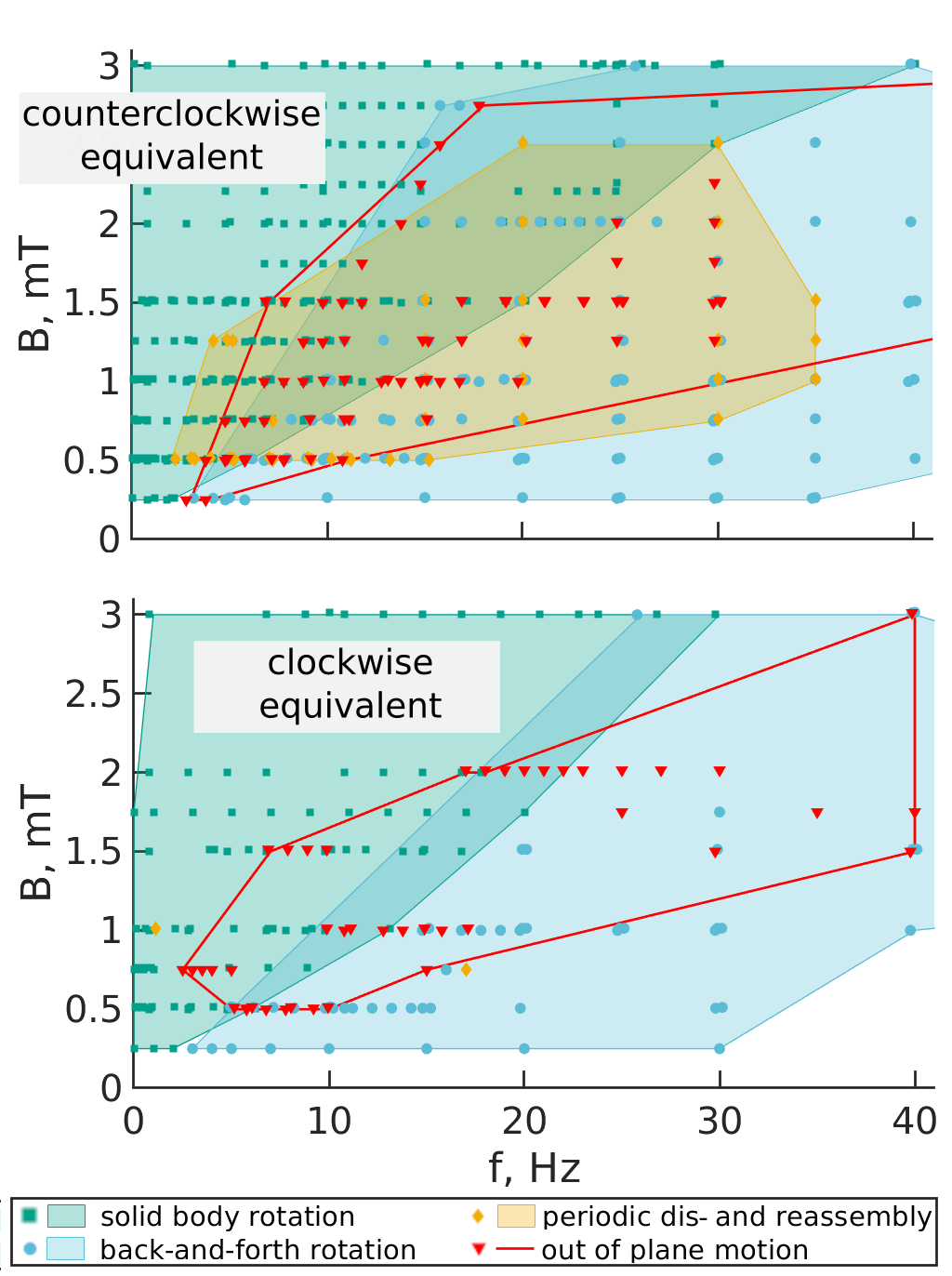}\\
    \caption{A phase diagram of the rotation regimes for the counterclockwise equivalent (top figure)  and clockwise  equivalent (bottom figure) orientation {with N=2 cubes per chain}. The areas corresponding to each rotation regime are defined by the extreme points of that regime.}
		\label{fig:phase_d_ex_towards}
\end{figure}

{ The data represented in Fig.\ref{fig:ffs} provides some insight into discrepancies between our theoretical understanding and experimental observations.} Data points clearly show that, although cube chains were selected as similar as visually possible, there are notable differences  in both characteristic values and rotation regimes. And it is not possible to explain them with clockwise equivalent ($+$ and $\times$) and counterclockwise equivalent (triangles) division. Moreover, chains not only have different critical frequencies at which they stop following the solid body rotation regime, but also follow different regimes at the same field and frequency values or change the regimes several times.

These observations hint that particle size differences { within the known limits for our sample}, surface effects and thermal fluctuations can cause significant differences in quantitative measurements among the different chains, which also goes to explain the discrepancies from theoretical predictions. { Calculations indicate that a difference in size of 10\% between cubes of the same chain would lead to considerable differences in the chain's behavior and critical frequency, compared to a chain of equally sized cubes. We know the distribution of cube sizes to exceed that and, while the cube pairs are selected on a basis of visual inspection to be as similar as possible, it is very likely that differences in size between cubes of the same dimer are in many cases significant enough to contribute to differences in dimer behavior, both comparing several dimers and experimental results to theoretical. Note that measurements of out of plane regime rotation frequencies should be considered less reliable than for the other regimes, due to difficulties of defining an angle when the chain is in near vertical position. Several points belonging to out of plane regime in Fig.~\ref{fig:ffs} coinciding with the curve of back-and-forth regime data points is a result of one dimer displaying behavior characteristic of both regimes at the same field magnitude and frequency. In such a case, the dimer would rotate as in back-and-forth regime but at a point transition into out of plane rotation. This transition could be driven by thermal fluctuations.}

{ Another reason for discrepancies between theory and experiment in Fig.~\ref{fig:ffs} is that mechanical friction between cubes is neglected in the theoretic model. When mechanical friction is taken into account, it is harder for cubes to slide along other cubes' faces.  This most  probably is the reason why in theoretical calculations the out of plane motion can be observed for higher frequencies than in the experiment.
}

As a result, the information from particle pair rotation experiments, which can be summarized in two phase diagrams, shown in Fig.~\ref{fig:phase_d_ex_towards}, involve overlapping areas. For better readability, as a frequency - field pair can correspond to several measurements, the data points have been offset. The areas corresponding to each rotation regime are defined by the extreme points of that regime.

The phase diagrams reveal that chain breakup  in examined magnetic fields  occurs predominantly when rotation happens in counterclockwise equivalent conditions, as would be expected from theoretical considerations (Fig.~\ref{fig:stability}).  
In experimental diagrams one observes that back-and-forth motion, particle disassembly and reassembly  and asynchronous out of plane rotation  slightly overlap with the solid-body rotation regime. {This overlap makes direct quantitative comparison between experimental phase diagrams and those seen in Fig.~\ref{fig:stability} impossible.}
While there is an overlap between all rotation regimes, the border between solid body and back-and-forth regimes are in the vicinity of those for entry into disassembly and reassembly  and asynchronous out of plane rotation motion regimes. In both orientations, a particle chain can only be reliably expected to remain in planar rotation and not enter 3D motion at high field - low frequency or, conversely, low field - high frequency conditions. 

\section{Conclusions}
\label{sec:concl}
In current work we examine a single cube and a short hematite chain dynamics in rotating magnetic fields. The investigation is mainly theoretical, but results for two-cube chains are verified also experimentally. To determine how important gravity effects are two models were developed, one with gravity and one without. 

For a single cube one finds that at low frequencies a cube rotates synchronously with the magnetic field and for higher frequencies asynchronous motion is observed. During synchronous motion one finds that a cube with rounded corners can rotate on the edge, corner or face. The magnetic moment is in the plane of the rotating magnetic field only if the cube rotates on an edge. 
Whether a cube rotates on an edge, corner and face depends on the magnetic field strength, frequency and initial conditions. In an asynchronous regime without explicit gravity effects there are two neutrally stable fixed points. Two modes of motion are observed: precession of magnetic moment and back-and forth motion. Including explicit gravity effects one finds that a new mode of complicated 3D motion appears. In this motion cube rotates slower than magnetic field, the lag increases, but instead of back motion to catch up with the magnetic field, the cube rolls and trough rotation in the third dimension catches up with the magnetic field. This mode has two sub-types where a cube rotates around one fixed point or around both of them.

For a two-cube system some of the motion modes which were observable for single cube disappear. This happens due to geometric restrictions.
No rotation on an edge and precession is possible. There appear, however, new scenarios: two-cube chains can break or asymmetric-chain is formed where cubes undergo different motion types, e.g., one cube rotates on a face while other on an edge. If the chain breaks then periodic chain  disassembly and reassembly is observed. For an individual chain dynamics depends on the clockwise or counterclockwise rotation direction of the magnetic field. However, when averaged over many chains, there is no dependence. The reason for this is that  in a large sample there are with the same probability two chain types. They behave differently at a given clockwise and counterclockwise rotation direction of the magnetic field.  But the first chain's type dynamics in a clockwise rotating magnetic field is  equal to the second chain's type dynamics in an counterclockwise rotating magnetic field. Thus, they balance out this effect and on average there are no differences.

In the case of small frequencies, the two-cube chain rotates synchronously with the magnetic field. Chain's configuration does not change for a fixed rotation frequency, thus the chain  rotates as a solid body. With increase of the frequency of the rotating magnetic field two scenarios can happen. Either chain breaks and enters to the mode where we observe periodic disassembly and reassembly of chain or asynchronous motion of chain is observed. 

The transition to the asynchronous regime for a two-cube system happens at lower rotational frequencies of the rotating magnetic field compared to a single cube. For asynchronous motion  depending on initial conditions three modes can be observed. One is the back-and-forth motion of the chain where the magnetic moment remains in the plane of the rotating magnetic field. The second is back-and-forth motion of the chain with periodical disassembly and reassembly. The third one is one of two modes: back-and-forth motion of asymmetric-chain or motion where the cube chain goes out of the plane of the rotating magnetic field.
There the chain, to catch up with the magnetic field, rolls on an edge and through rotation in the third dimension  catches up with the magnetic field. The last mode to our knowledge has not been described before in scientific literature. 

For experimental conditions ($B_{exp}\in [0.3; 3]\,\mathrm{ mT}$ and $a_0\approx1.5\,\mathrm{\mu m}$) for a two-cube chain  four regimes of motion are possible. Three of them are planar motion regimes: solid-body motion, back-and-forth motion, and periodic chain disassembly and reassembly, and motion where the cube chain goes out of the plane of the rotating magnetic field. All four modes were observed, identified in our experiments. Also there was no indication that there should exist another mode. In experiments, as predicted in the theory part of this paper, one observes that there are two chain types for which motion differs for a given clockwise or counterclockwise rotation of a magnetic field. But the motion dynamics for the first chain type in case of a clockwise rotation of the magnetic field is equal with the second chain type's dynamics in counterclockwise rotation. Thus, in a large sample there is no global dependence on rotation direction as in equilibrium there is an equal number of particles in each configuration. Theoretical consideration predicts that an applied field rotating in one direction could change the balance between configurations, however, no experimental  indications of that are observed.

\section*{Acknowledgment}

M.B. acknowledges financial support from PostDocLatvia grant No. 1.1.1.2/VIAA/3/19/562.

\bibliography{biblography}

\end{document}